\documentstyle[aps,multicol,prb,epsf]{revtex}

\begin{document}
\draft 

\title{Subgap transport in ferromagnet-superconductor junctions
due to magnon-assisted Andreev reflection}

\author{Grygoriy Tkachov$\sp{1,2}$, Edward McCann$\sp{1}$ and
Vladimir I. Fal'ko$\sp{1}$}
\address{$\sp{1}$Department of Physics, Lancaster University,
Lancaster, LA1 4YB, United Kingdom}
\address{$\sp{2}$Institute for Radiophysics and Electronics NAS, Kharkiv,
12 Acad. Proscura St., 61085, Ukraine}
\date{\today}
\maketitle \begin{abstract} {
We propose a new process of magnon-assisted Andreev reflection
at a ferromagnetic metal - superconductor interface,
which consists of the simultaneous injection of a Cooper pair from the
superconductor and the emission of a magnon inside the ferromagnet.
At low temperature this process represents an additional channel for sub-gap
transport across an FS interface, which lifts restrictions on the current
resulting from the necessity to match spin-polarized current in the ferromagnet
with spin-less current in the superconductor.
For a junction between a superconductor 
and a ferromagnet with an arbitrary degree of polarization,
the inelastic magnon-assisted Andreev reflection process
would manifest itself as a nonlinear addition to the $I(V)$ characteristics
which is asymmetric with respect to the sign of the bias voltage and
is related to the density of states of magnons in the ferromagnet.
Expressions for the subgap $I(V)$ characteristics are given for arbitrary
interfacial quality whilst the limiting cases of uniformly transparent
and disordered interfaces are discussed in detail.
}\end{abstract}

\begin{multicols}{2}

\bibliographystyle{simpl1}
%
%
\section{Introduction}
\label{intro}


Spin polarized transport is a subject of intense research,
motivated by the desire to develop a form of electronics which utilizes the
spin polarization of carriers\cite{pri95}.
Ferromagnetic (F) metals have more carriers of one spin polarization
(known as majority carriers) present at the Fermi energy $E_F$
than of the inverse polarization (minority carriers) and,
in a ferromagnet-ferromagnet (FF) junction, the resistance depends
on the relative orientation of the magnetization in the ferromagnets.
With parallel magnetizations, carriers flow from majority
to majority bands (and minority to minority) whereas in a junction with
antiparallel magnetizations,
carriers flow from majority to minority bands (and vice versa).
The resulting spin current mismatch produces a larger contact resistance
in the antiparallel orientation, an effect known as tunneling
magnetoresistance in FF junctions\cite{jul75,moo00}
and giant magnetoresistance in multilayer structures\cite{bai88,pra91}.

Meanwhile, recent achievements in fabrication techniques have led to the
possibility of creating hybrid heterostructures combining ferromagnetic (F)
and superconducting (S) elements and
it has been pointed out\cite{T+M71,deJ95,Fal99a} that the effect of
spin current mismatch may affect the conductance of a
ferromagnet-superconductor junction\cite{vas97,sou98,pet99,L+G99,Gir98,Upa98}.
At low temperatures and small bias voltage
(subgap regime, $T\ll \Delta $, $eV<\Delta $), current flows
through the interface due to Andreev reflection\cite{And64}
whereby particles in the ferromagnetic region with excitation energies
$\epsilon$ smaller than the superconducting gap energy $\Delta$
are reflected from the interface as holes.
Since subgap transport in the superconductor (S) is mediated by
spinless Cooper pairs the spin current is zero in the
superconductor in contrast to the ferromagnet.
In high-quality metallic type
FS junctions, this has been shown to result in a non-equilibrium spin
accumulation in the vicinity of the FS interface \cite{Fal99a,Jed99,Fal99b}.
Such an accumulation of non-equilibrium spin density generates a
compensating spin-current flow and, therefore, an additional non-local
interfacial resistance\cite{deJ95} formed at the spin-relaxation length
scale in a ferromagnet\cite{Fal99a,Fal99b}. 

An increasing number of experiments have been devoted to studying the transport
properties of ferromagnet/insulator/superconductor junctions. 
Some\cite{T+M71,T+M+mons} have measured the degree of spin-polarization of various 
ferromagnets and others\cite{vas97,wei99} have studied the effect of spin-polarized
quasiparticle injection into high-$T_c$ superconductors.
In tunnel FS junctions, the necessity to match microscopic spin currents at the interface 
creates an opportunity for complex electron transfer processes to manifest themselves
in the current formation. For instance, 
spin relaxation phenomena\cite{pri95,j+s85}, such as spin-orbit 
scattering at impurities or magnon emission, can reduce the spin current mismatch.
In a given junction, spin-orbit scattering (which is an elastic process)
would cause a linear reduction in the value of the additional
contact resistance\cite{Fal99b,Bax99}. On the other hand the inelastic process of
magnon emission would manifest itself as a modification of the
form of the $I(V)$ characteristics. Indeed, nonlinear $I(V)$ characteristics due to
magnon-assisted tunneling between two ferromagnets have already been studied both 
theoretically\cite{bra98,mac98,gui97} and experimentally\cite{tsu71,moo98}
with a view to relate the second derivative of the current
to the density of states of magnons $\Omega (\omega)$
as $d^2I/dV^2 \propto \Omega (eV)$. 

The aim of this paper is to investigate Andreev reflection accompanied by the emission 
of magnons in tunnel contacts between a ferromagnet and a superconductor (a preliminary
discussion of the half-metallic ferromagnet-superconductor case is given in Ref.\onlinecite{m+f}). 
For this purpose we calculate the nonlinear subgap current $I$ using the tunneling Hamiltonian 
method\cite{bar61} and the nonequilibrium Green functions 
technique\cite{Caroli,Cuevas,Keldysh} generalized for describing contacts with ferromagnetic 
electrodes. The differential conductance $G(V)=dI/dV$ ($V$ is the voltage) 
is shown to be the sum of the usual Andreev conductance $G_A$ and a contribution $G_{in}$ 
from inelastic processes in the ferromagnet: $G=G_A+G_{in}$. For a uniform partially transparent 
interface $G_A$ is given at low temperatures by\cite{btk,deJ95}

\begin{equation}
G_A(V)
=\frac{4e^2}{h}|t|^4\frac{A\Pi_{\downarrow}}{h^2(1-(eV/\Delta)^2)},\quad |t|^4\ll 1,
\label{GA}
\end{equation}
%
where the factor $|t|^4$ stands for the probability of the simultaneous 
tunneling of two particles involved in the process of Andreev reflection,
$h$ is the Planck constant and $e$ is the absolute value of the elementary charge.
Throughout this paper we use the notation $\uparrow$ and $\downarrow$ to represent
majority and minority spin bands, respectively, and the factor $\Pi_{\downarrow}$
in Eq.~(\ref{GA}) denotes the area of the maximal cross section of the Fermi surface of minority (spin 
`down') electrons in the plane parallel to the interface, so that Andreev conductance is proportional to the number of
minority spin channels $A\Pi_{\downarrow}/h^2$ in a contact with area $A$.

The inelastic contribution $G_{in}$ to the differential conductance results from the interaction between spins of 
the conduction electrons and the quasilocalized moments of the inner atomic shells which are responsible for the 
magnetism. In the second quantization language emerging from the Holstein-Primakoff transformation\cite{HP} 
this interaction -- treated within the s-f (s-d) 
model\cite{FCo,Nolting} -- represents magnon emission or absorption accompanied by a spin-flip process which is 
preceded or followed by Andreev reflection. In the zero-temperature limit where the magnon emission is only 
allowed the inelastic conductance $G_{in}$ is given for a uniform interface by

\begin{eqnarray}
G_{in}(V) &=& \frac{e^2}{hS}|t|^4
\left[
M(eV)
\frac{A\Pi_{\uparrow}}{h^2}
+
M(-eV)
\frac{A\Pi_{\downarrow}}{h^2}
\right],
\label{Gin}\\
M(x) &=& \Delta^2\Theta(x)
\int_{0}^{2x}\Omega(\omega)d\omega
\times\label{M}\\
&& \qquad \quad
\times\left[
\frac{1}{\Delta^2-x^2}+
\frac{1}{\Delta^2-(x-\omega)^2}
\right],
\nonumber
\end{eqnarray}
%
$$
2eV<2\Delta\ll E_S.
$$
The first term in Eq.~(\ref{Gin}) represents the contribution from the majority (spin `up') states close to the Fermi
level and $\Pi_{\uparrow}$ denotes the maximal cross section of the majority Fermi surface in the plane parallel 
to the interface. The second contribution is related to electronic states located in the vicinity of the minority 
Fermi surface. It is assumed that the magnon local density of states near the interface, $\Omega(\omega)$ in Eq.~(\ref{M}),
is non-zero for $\omega\geq\omega_0$, where $\omega_0$ is a relativistic anisotropy gap in the magnon spectrum which 
may arise from a spin-orbit coupling or from the demagnetization geometry of an F-layer (see, e.g., Ref~\onlinecite{Levy}). 
$S$ is the spin of the localized moments.
In Section~\ref{Sub} we describe in more detail the model and technique used, 
and we give general results for the $I(V)$ characteristics valid for arbitrary interfacial quality.

Both terms in the square brackets in Eq.~(\ref{Gin}) are asymmetric functions of voltage (this is emphasized by 
the presence of the unit step function $\Theta$ in Eq.~(\ref{M})). 
The first term is non-zero for positive voltage $V>\omega _{0}/(2e)$, while the second one is non-zero
for negative voltage $V<-\omega _{0}/(2e)$.  This feature can be explained using the sketch in
Fig.~1 which illustrates the tunneling process between an S electrode on the
left hand side and an F electrode on the right for (a) $V>0$ and (b) $V<0$.
We have adopted the convention that positive (negative) voltage results in a
Fermi energy $E_{F}$ in the ferromagnet that is lower (higher) by energy
$|eV|$ than the Fermi energy $E_{S}$ in the superconductor. For $V>0$,
Fig.~1(a), Andreev reflection results in the injection of both a majority
(spin `up') and a minority (spin `down') electron into the ferromagnet. The
electron energies, $\pm \epsilon $ with respect to $E_{S}$ and $eV\pm
\epsilon $ with respect to $E_{F},$ are both above $E_{F}$ because of the
need to move into unoccupied states in the ferromagnet, $\epsilon \leq eV$.
Due to the existence of the intra-atomic exchange interaction with the 
localized moments, a dynamic process, shown
schematically in Fig.~1(a), allows a spin down electron to tunnel into a
virtual, intermediate spin down state above $E_{F}$ and then emit a magnon,
depicted in Fig.~1(a) as a flip of a core spin, enabling the electron to
incorporate itself into an empty state in the majority, spin up conduction band.
Since the magnon carries away spin equal to -1, the total spin is equal to zero in the above 
process. The relevant magnon energies $\omega$ are less than the energy of a Cooper pair $2eV$ 
measured with respect to the chemical potential in the ferromagnet (see Eq.~(\ref{M})). Since 
the superconducting gap energy $\Delta$ is usually much less than the magnon Debye energy, 
the inequalities $\omega\leq 2eV\leq 2\Delta$ imply that mainly long-wavelength magnons with
momenta much smaller than the Fermi momentum assist the subgap transport. This means in turn
that only electron states near the majority Fermi surface are involved in the magnon-assisted
transport at $V>0$. 

On the other hand, for $V<0$, Fig.~1(b), Andreev reflection leads to 
the injection of a spin up and a spin down electron from the ferromagnet
into the superconductor with energies $\epsilon $ above and below $E_{S}$.
Due to the intra-atomic exchange the following dynamic process can contribute
to current formation at zero temperature $T=0$: one of the spin down electrons below
$E_F$ emits a long-wavelength magnon and forms a virtual, intermediate spin up state. Such a process 
allows it to tunnel into the superconductor along with another spin down carrier. Their energies, 
$-|eV|\pm \epsilon$ with respect to $E_{F}$, are both below $E_{F}$ because of the need to 
have initially occupied states in the ferromagnet. Since emission of a long-wavelength 
magnon does not change significantly the energy and momentum of the spin down electron, the initial and 
intermediate electron states are close to the minority Fermi surface. 
We will point out later that both the majority and minority magnon-assisted currents are quadratic 
in the interaction constant $\Gamma$ relevant to the intra-atomic exchange. On the other hand, they are also 
quadratic with respect to the electron lifetime in the virtual state which is of the order of $\hbar/\Gamma$. 
For this reason Eq.~(\ref{Gin}) for the magnon-assisted conductance does not contain the interaction 
constant. 

%
\begin{figure}[t]
\epsfxsize=1.0\hsize
\epsffile{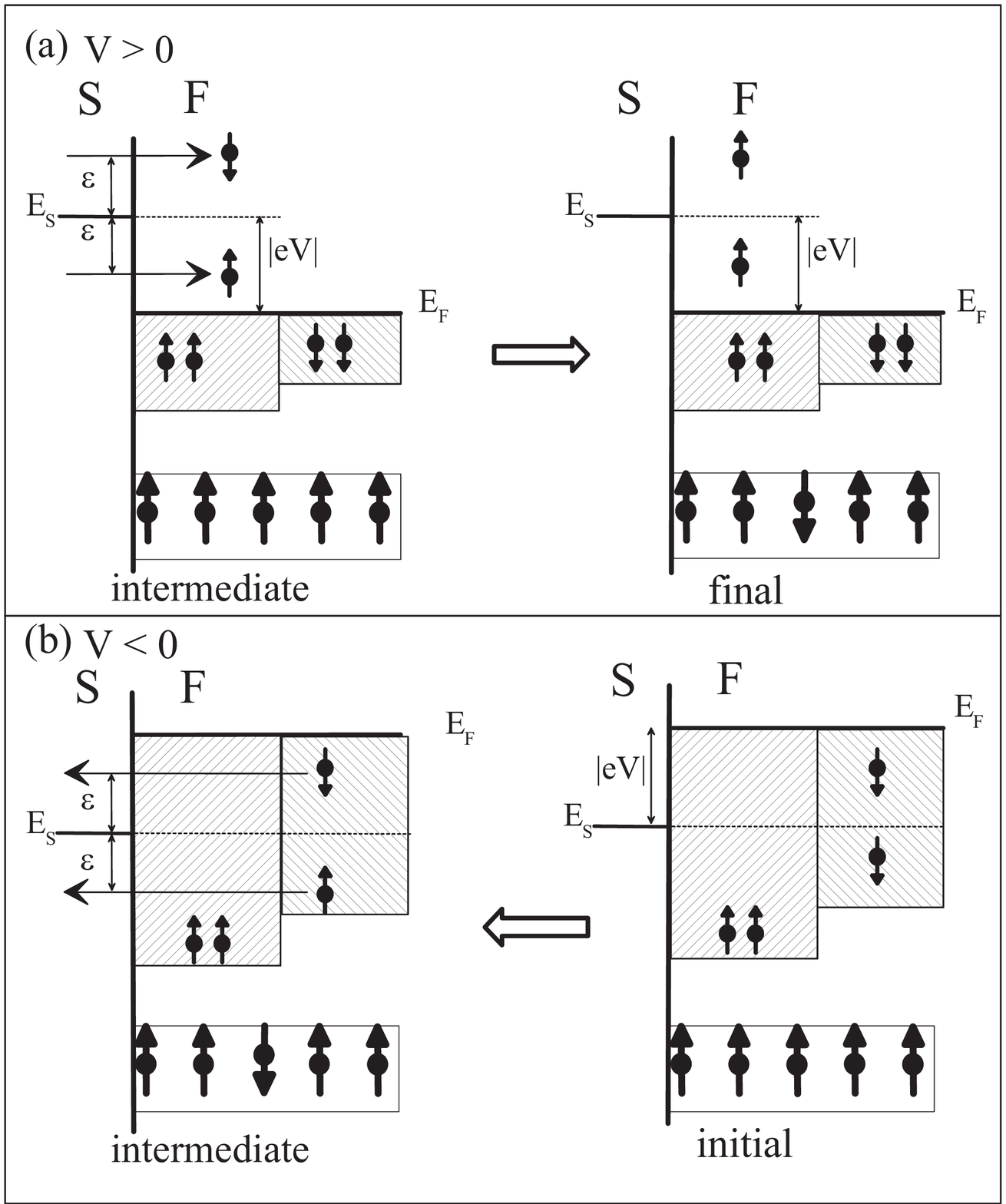}
\refstepcounter{figure}
\label{figure:1}

{\setlength{\baselineskip}{10pt} FIG.\ 1 - Schematic representation of the 
magnon-assisted tunneling process between a superconductor on the
left hand side and a conventional ferromagnet on the right for
(a) $V>0$ and (b) $V<0$.
For (a) $V>0$ a spin down electron tunneling into the ferromagnet may emit
a magnon and incorporate itself into the majority (spin up) conduction band.
For (b) $V<0$ a spin down electron with energy below $E_F$ may emit a magnon
and form an intermediate spin up state before tunneling into the
superconductor.
}
\end{figure}
%

We note that the total conductance $G=G_A+G_{in}$ in Eqs. (\ref{GA}) -- (\ref{M})
is also an asymmetric function of voltage due to the inelastic processes in the ferromagnet. This asymmetry is related 
to the difference $\Pi_{\uparrow}-\Pi_{\downarrow}$ in the maximal cross sections of the majority and minority Fermi 
surfaces and can be used to extract the magnon-assisted (inelastic) contribution from transport measurements.
The asymmetric contribution to the conductance may be related
 to the magnon density of states by taking the derivative
with respect to voltage of the difference $G(V)-G(-V)$: 
\begin{eqnarray}
&&\frac{d}{dV}[G(V)-G(-V)] = \nonumber \\
&& \quad =  \, \frac{4e^3 A |t|^4}{h^3 S}
\left[ \Pi_{\uparrow} - \Pi_{\downarrow} \right]
\Biggl\{ \frac{\Omega (|2eV|)}
{\left[1-\left( |eV| /\Delta \right)^{2}\right]}
+ {} \nonumber \\
&& \qquad {} + \int_{0}^{|2eV|} \Omega (\epsilon )
\left[ F(\frac{|eV|}{\Delta })-F(\frac{\epsilon - |eV|}{
\Delta })\right] \frac{d\epsilon }{2\Delta }\Biggr\} , 
\label{iv2}
\end{eqnarray}
%
where $F(x) = x/(1-x^2)^2$.
An extreme example occurs in tunneling involving a half-metallic ferromagnet which is a material where the
exchange spin splitting between the majority and minority spin bands exceeds the Fermi energy, measured from the
bottom of the majority band, so there are only majority spin carriers at the Fermi energy\cite{g+s}.
Since $\Pi_{\downarrow} = 0$,
the contributions of conventional Andreev reflection, Eq.~(\ref{GA}), and of states near the minority Fermi surface
to the inelastic process, second term in Eq.~(\ref{Gin}), are absent and the contribution of majority states
to the inelastic process, first term in Eq.~(\ref{Gin}), dominates the subgap $I(V)$ characteristics\cite{m+f}.
At the beginning of Section~\ref{Cond} we discuss junctions
involving a half-metallic ferromagnet in more detail
and we present results (Eq.~(\ref{GinHFS})) for the differential conductance
of both uniformly transparent and disordered planar interfaces.

Since the inelastic conductance (\ref{Gin}), (\ref{M})
is written in terms of the local density of states 
$\Omega(\omega)$ of long-wavelength magnons at the
interface, the same expressions may be used to describe
ferromagnetic systems with more complicated magnon spectra.
For example, a ferromagnetic superlattice or a trilayer
(see, e.g., Ref.~\onlinecite{Stiles} and Refs therein)
can be used as an F element of the contact.
A suggested geometry, sketched in Fig.~2, consists of two
superconducting reservoirs which sandwich a
ferromagnetic-normal metal multilayer adjacent to a `long' normal layer.
The `long' normal layer produces asymmetry in the structure
(e.g. by introducing an additional resistance)
so that the inelastic contribution to the
conductance would depend on the local density of states
of magnons at the interface between the
multilayer and the superconducting reservoir on the left.
Note that the use of superconducting reservoirs is common
when measuring the current perpendicular to plane (CPP)
giant magnetoresistance of multilayer structures
because they reduce the overall resistance of the system.
In multilayer structures the spectrum of magnons 
may consist of several separate bands\cite{MSpectrum}
so in contacts where the superconducting
gap energy $\Delta$ is larger than the typical magnon
bandwidth the magnon band structure may manifest itself
in the form of the $I(V)$ characteristics.
The precise form and the influence of the magnon spectrum in the formation
of the subgap current should depend on whether the coupling 
between magnetic layers is ferromagnetic or antiferromagnetic.   

The paper is organized as follows. In Section \ref{Sub} we introduce the model and technique used for describing a tunnel 
FS contact and give general results for the subgap current and the $I(V)$ characteristics valid for an arbitrary interfacial
quality. In Section \ref{Cond} we calculate the differential conductance for both a half-metallic 
ferromagnetic electrode and a conventional ferromagnet. The results for uniform and disordered interfaces are discussed. 
Appendix \ref{Form} is devoted to detailed calculations of the subgap current based on the Keldysh formalism. 

%
\begin{figure}[t]
\hspace{0.06\hsize}
\epsfxsize=0.8\hsize
\epsffile{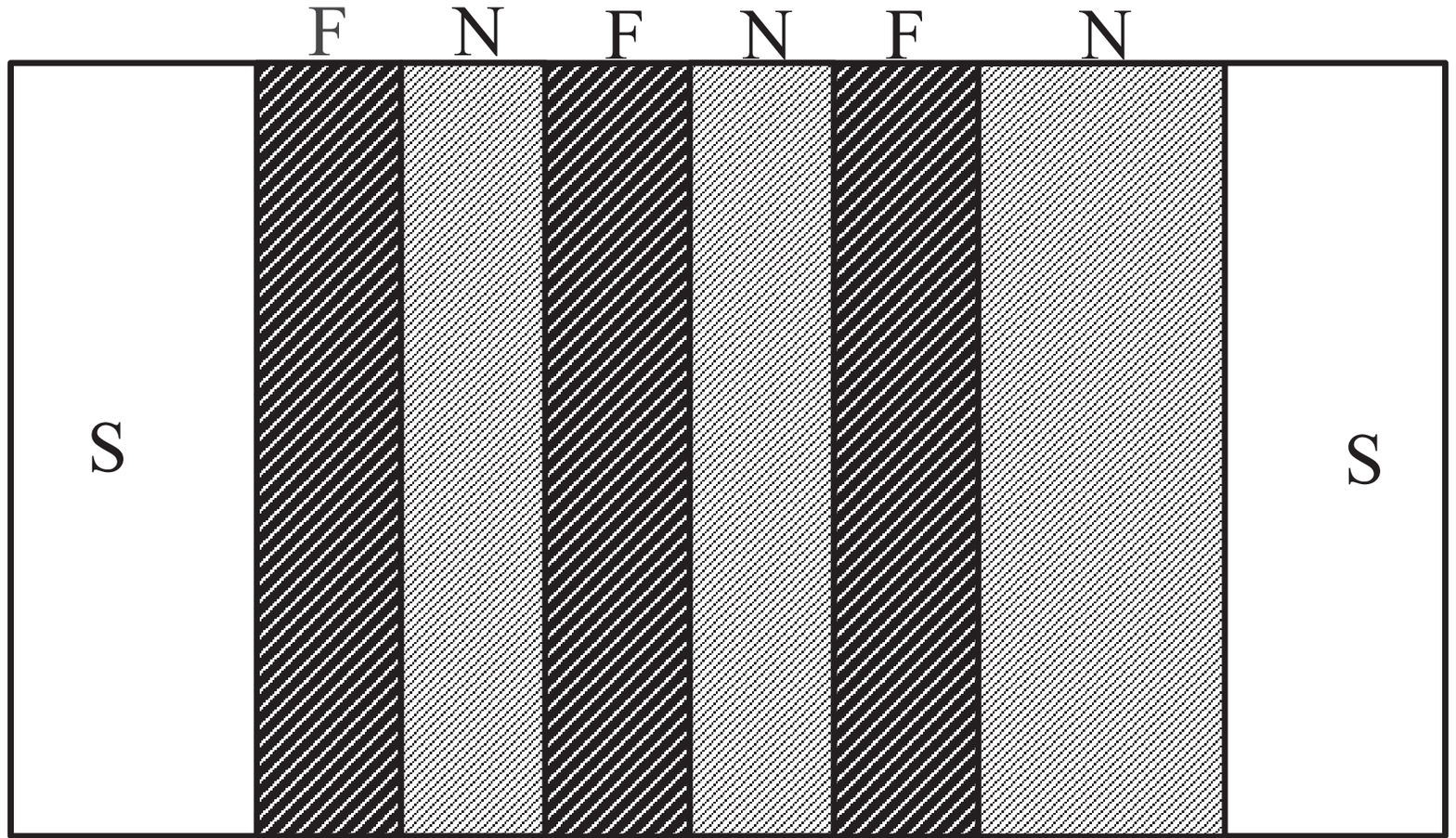}
\vspace{0.1cm}
\refstepcounter{figure}
\label{figure:2}

{\setlength{\baselineskip}{10pt} FIG.\ 2 - Sketch of a suggested
geometry in which the contribution of magnon-assisted Andreev reflection 
to the differential conductance is related to the magnon density of states
of a magnetic multilayer system.
Two superconducting reservoirs (S) sandwich a ferromagnetic-normal metal
multilayer (F/N) adjacent to a `long' normal layer (N).
}
\end{figure}
%

\section{Subgap transport through a ferromagnet-superconductor interface}
\label{Sub}
%
\subsection{Model}

In what follows we consider a tunnel contact of area $A$ between a ferromagnet and a superconductor
which have dimensions $L_F$ and $L_S$, respectively, in the direction $z$ perpendicular to the contact 
plane. The contact is positioned at $z=0$ so that $-L_S\leq z\leq L_F$.
The difference $E_S-E_F=eV$ in the chemical potentials of the electrodes is 
assumed to be smaller than the superconducting gap energy $\Delta$:  $|eV|<\Delta$ and the temperature is 
finite, but also less than the gap energy: $T<\Delta$. 

To describe electronic tunneling through the barrier at the interface ($z=0$) we make use of the tunneling 
Hamiltonian method\cite{bar61} and the nonequilibrium Green functions technique \cite{Caroli,Cuevas,Keldysh}. 
In this approach the total Hamiltonian of the system is

\begin{eqnarray}
&
H=H_S+H_F+H_T,
&\label{H}\\
& 
H_T=\sum_{kk^\prime \alpha}\left[
t_{SF}({\bf k}^\prime, {\bf k})c^\dag_{k^\prime\alpha}a_{k\alpha}+\mbox{h.c.}
\right].
&
\label{HT}
\end{eqnarray}
where $H_T$ is the tunneling Hamiltonian,
$H_F(a^\dag_{k\alpha},a_{k\alpha})$ is the Hamiltonian of the ferromagnetic electrode,
and $H_S(c^\dag_{k^\prime\alpha},c_{k^\prime\alpha})$ is the Bogolubov de Gennes Hamiltonian of a superconductor (hereafter, 
electron wave vectors in the S side are marked with prime). The so 
called s-f (s-d) model \cite{FCo,Nolting}, which assumes that magnetism and electrical conduction are caused by different
groups of electrons, is employed to describe the ferromagnetic electrode. The magnetism originates from the inner atomic 
shells (e.g., d or f) which have unoccupied electronic orbitals and, therefore, possess magnetic moments. On the other hand, the 
conduction is related to electrons with delocalized wave functions. These two electron groups are considered to 
be coupled via an intra-atomic exchange interaction. Using the Holstein-Primakoff transformation \cite{HP} the 
operators of the localized moments in the interaction Hamiltonian can be expressed via magnon creation and 
annihilation  operators $b^\dag,  b$.
At low temperatures, where the average number of magnons is small $<b^\dag  b>\ll 2S$ ($S$ is the spin of the localized moments),
the Hamiltonian of the ferromagnet $H_F$ can be written as follows

\begin{equation}
H_F=H_e+H_m+H_{em},
\label{HFsw}
\end{equation}
\begin{eqnarray}
H_e=\sum_{k\alpha}\epsilon_{\alpha}({\bf k})a^\dag_{k\alpha}a_{k\alpha},
&\quad& 
\epsilon_{\alpha}({\bf k})=
\epsilon_F({\bf k})-\alpha\Gamma\gamma({\bf k})/2,
\nonumber\\
\alpha=\pm &\quad& \mbox{or}\quad \uparrow,\downarrow,
\label{He}
\end{eqnarray}
\begin{equation}
H_m=\sum_q\omega_qb^\dag_qb_q,\quad \omega_{q=0}=\omega_0,
\label{Hm}
\end{equation}
\begin{eqnarray}
&&
H_{em}=-\frac{1}{(2S)^{1/2}}\sum_{qkp}\left[
U_{qkp}b^\dag_q a^\dag_{k\uparrow}a_{p\downarrow}+
\mbox{h.c.}
\right],
\nonumber\\
&&
U_{qkp}=\Gamma u({\bf k},{\bf p})(AL_F)^{1/2}\int d{\bf r}
\chi^*_q({\bf r})\Phi^*_k({\bf r})\Phi_p({\bf r}).
\label{Hem}
\end{eqnarray}
The first term $H_e$ in (\ref{HFsw}) deals with conduction band electrons which are split into majority 
$\epsilon_{\uparrow}({\bf k})$ and minority $\epsilon_{\downarrow}({\bf k})$ subbands due to the s-f (s-d) 
exchange, $\epsilon_F({\bf k})$ is the bare electron energy, $\Gamma$ sets the scale for the splitting energy and 
the dimensionless function $\gamma({\bf k})$ takes into account a possible anisotropy of the band splitting. The 
Hamiltonian $H_m$ (\ref{Hm}) describes free magnons with spectrum $\omega_q$ which in the general case has 
a gap $\omega_{q=0}=\omega_0$. The third term $H_{em}$ in (\ref{HFsw}) is the electron-magnon coupling (\ref{Hem}) 
resulting from the intra-atomic exchange interaction between the spins of the conduction electrons and the localized 
moments. The matrix element of this interaction $U_{qkp}$ is, generally speaking, of the same order 
as the band splitting. In the expression for $U_{qkp}$ (see Eq.~(\ref{Hem})) $\Phi_k({\bf r})$ and $\chi_q({\bf r})$ are 
the electron and magnon wave functions, respectively, which are normalized by the volume $AL_F$ of the ferromagnet. 
The dimensionless function $u({\bf k},{\bf p})$ introduces an extra dependence 
on the electron wave vectors which may come from an anisotropy of the electron-magnon coupling near the Fermi 
surfaces of the majority and the minority electrons. 
 
The tunneling Hamiltonian $H_T$ (\ref{HT}), which couples the two electrodes, is written in terms
of the creation and annihilation Fermi operators $c^\dag,c$  and  $a^\dag,a$ of the S and F electrodes respectively. 
The tunneling matrix elements $t_{SF}({\bf k^\prime},{\bf k})$ describe the transfer of an electron with wave 
vector ${\bf k}$ from the ferromagnet (F) to the state with ${\bf k}^\prime$ in the superconductor (S). We will consider 
$t_{SF}({\bf k^\prime},{\bf k})$ to be a symmetric matrix of the form

\begin{eqnarray}
t_{SF}({\bf k^\prime},{\bf k}) &=& t_{{\bf k^\prime}_{||}, {\bf k}_{||}}
\left|\frac{h^2  v_S^z({\bf k}^\prime_{||},k^{\prime}_z)  v_F^z({\bf k}_{||},k_z)}{2L_SL_F}\right|^{1/2} \!,
\label{t}\\
\end{eqnarray}
where
\begin{eqnarray}
v_{S,F}^z({\bf k}_{||},k_z) &=&
\partial\epsilon_{S,F}({\bf k})/\partial(\hbar k_z).
\nonumber
\end{eqnarray}
This expression explicitly takes into account the fact that the 
quantum flows $v_{S,F}^z/L_{S,F}$ perpendicular to the interface 
are non-zero on either side of the junction for electrons participating in the tunneling process; $v_{S,F}$ are 
the perpendicular velocities and $\epsilon_{S,F}({\bf k})$ denote the electron spectra in the 
superconductor and the ferromagnet.
The wavenumbers ${\bf k}$ are separated into components perpendicular $k_z$ and parallel ${\bf k}_{||}$ to
the interface, so that the dimensionless factor $t_{{\bf k^\prime}_{||}, {\bf k}_{||}}$ depends only on the momenta 
parallel to the interface. We also suppose that spin is conserved upon electronic transfer across the interface
and pay attention to the spin index $\alpha=\{\uparrow,\downarrow\}$ (or $\pm$) in all calculations. 

In what follows it is assumed that the voltage drop $V$ occurs across the junction so that both electrodes are almost 
in  equilibrium. The total current $I$ through the interface is  proportional to the average value of the rate of 
change of the number of particles  in, for  example,  the  ferromagnet:  $I=e<\dot{N}_F>$.  This rate can be found 
from the commutator of  $N_F=\sum_{k\alpha}a^\dag_{k\alpha}a_{k\alpha}$  with the total  Hamiltonian  
$H=H_S+H_F+H_T$.  Bearing in mind that only the tunneling  Hamiltonian $H_T$ fails to commute with $N_F$, we 
arrive at the following formula for $I$

\begin{eqnarray}
I &=& \frac{ie}{\hbar}\sum_{kk^\prime \alpha}
\left[
t_{SF}({\bf k}^\prime, {\bf k})<c^\dag_{k^\prime \alpha}a_{k\alpha}>-
\right.
\label{I}\\
&& \qquad
\left.
-t_{SF}^*({\bf k}^\prime, {\bf k})<a^\dag_{k\alpha}c_{k^\prime \alpha}>
\right].
\nonumber
\end{eqnarray}
The averaged  quantities in Eq.~(\ref{I}) can be expressed in terms of nonequilibrium (Keldysh) Green functions
${\cal G}^{+-}(t,t)$ and ${\cal G}^{-+}(t,t)$ \cite{Caroli,Cuevas,Keldysh} and it is convenient for the case of a 
superconducting electrode to use the $2\times 2$ Nambu representation
as shown explicitly in Eq.~(\ref{G}) of Appendix \ref{Form}. The 
anti-hermitian relationships Eq.~(\ref{anti}) for ${\cal G}^{+-}$ and ${\cal G}^{-+}$ enable us to 
represent the current $I$ (\ref{I}) in terms of two off-diagonal
(with respect to the electrode subscripts $\{ F,S \}$) Keldysh functions:

\begin{eqnarray}
I &=& 2e\, Re\sum_{kk^\prime}\left[
t_{SF}({\bf k}^\prime, {\bf k}){\cal G}^{+-}_{FS|11}({\bf k}t,{\bf k^\prime}t) +
\right.
\label{IKeld}\\
&& \qquad
\left.
+t_{SF}^*({\bf k}^\prime, {\bf k}){\cal G}^{-+}_{FS|22}({\bf k}t,{\bf k^\prime}t)
\right], 
\nonumber 
\end{eqnarray}
where the subscripts $\{ 1,2 \}$ refer to spin indices and the symbol $Re$ means the real part of the sum.

\subsection{Andreev and magnon-assisted currents. Derivation of the subgap $I(V)$ characteristics}

The Green functions ${\cal G}^{+-}_{FS|11}$ and ${\cal G}^{-+}_{FS|22}$ in (\ref{IKeld}) can be determined with the 
help of perturbation theory. For this purpose, we represent the total Hamiltonian (\ref{H}) in the form 
$H=H_S+H_e+H_m+(H_T+H_{em})$ where the sum of the tunneling Hamiltonian and the Hamiltonian of the 
electron-magnon interaction is treated as a perturbation. In this approach the usual Andreev corrections to 
Green functions ${\cal G}^{+-}_{FS|11}$ and ${\cal G}^{-+}_{FS|22}$ appear in the third order of the perturbation
with respect to the tunneling (and zeroth order in the electron-magnon interaction) and are proportional to

\begin{equation}
\int_{K}d\tau_1d\tau_2d\tau_3H_T(\tau_1)H_T(\tau_2)H_T(\tau_3),
\label{APert}
\end{equation}
where the  integration  is  performed  along  the  Keldysh  contour (see Eq.~(\ref{IntKeld}) in Appendix \ref{Form}). 
They give rise to a tunneling current $I$ (\ref{IKeld}) which is of the fourth order in the hopping elements 
$t_{k^\prime,k}$ as it should be because two particles are involved in the process of Andreev reflection. 
The lowest order contribution to magnon-assisted Andreev reflection also arises in the third order of the
tunneling matrix  $t_{ k^\prime,k}$, but it is bilinear in the electron-magnon interaction,
\begin{eqnarray}
&
\int_{K}d\tau_1d\tau_2d\tau_3d\tau_4d\tau_5\times
&\label{MAPert}\\ 
&
\times
H_T(\tau_1)H_T(\tau_2)H_T(\tau_3)H_{em}(\tau_4)H_{em}(\tau_5).
&\nonumber
\end{eqnarray} 
Since the interaction  Hamiltonian $H_{em}$ (\ref{Hem}) is of the first order in the 
magnon operators $b^\dag_q$, $b_q$, the diagrams corresponding to this term will contain only one magnon line. The 
diagrammatic calculation of the tunneling current (\ref{IKeld}) performed in Appendix \ref{Form} show that $I$ can be 
represented as the sum of three different contributions:

\begin{equation}
I=I_{\uparrow\downarrow}+I_{\uparrow\uparrow}+I_{\downarrow\downarrow},
\label{ISub}
\end{equation}
\end{multicols}\vspace*{-3.5ex}{\tiny
                   \noindent\begin{tabular}[t]{c|}
                   \parbox{0.493\hsize}{~} \\ \hline \end{tabular}}
\begin{eqnarray}
&&
I_{\uparrow\downarrow}=\frac{4e}{\hbar}\int d\epsilon\sum_{k1,k2}
\left|\sum_{k^\prime}
t_{SF}({\bf k}^\prime,{\bf k}_1)t_{SF}(-{\bf k}^\prime,{\bf k}_2)g_{SS|12}({\bf k^\prime})\right|^2
\delta(-\epsilon+eV-\xi_{k1,\uparrow})
\delta(2eV-\xi_{k1,\uparrow}-\xi_{k2,\downarrow})
\times
\label{Iud3}\\
&&
\qquad \qquad \qquad \qquad \qquad \qquad \qquad \qquad \qquad \qquad \qquad \qquad \qquad \times
\left\{
n(-\xi_{k1,\uparrow})n(-\xi_{k2,\downarrow})-
n(\xi_{k1,\uparrow})n(\xi_{k2,\downarrow})
\right\},
\nonumber
\end{eqnarray}
%
\begin{eqnarray}
&&
I_{\uparrow\uparrow}=\frac{2e}{\hbar S}
\int d\epsilon d\omega\sum_{q,k1,k2}
\left|\sum_{k^\prime}
t_{SF}({\bf k}^\prime,{\bf k}_1)t_{SF}(-{\bf k}^\prime,{\bf k}_2)g_{SS|12}({\bf k^\prime})\right|^2
\left|\sum_p\frac{U_{qk2p}}{\xi_{p,\downarrow}-\xi_{k2,\uparrow}-\omega}\right|^2 
\times\label{Iuu3}\\
&&
\times
\delta(\omega-\omega_q)
\delta(-\epsilon+eV-\xi_{k1,\uparrow})
\delta(2eV-\xi_{k1,\uparrow}-\xi_{k2,\uparrow}-\omega)
\left\{
n(-\xi_{k1,\uparrow})n(-\xi_{k2,\uparrow})[1+N(\omega)]
-
n(\xi_{k1,\uparrow})n(\xi_{k2,\uparrow})N(\omega)
\right\},
\nonumber
\end{eqnarray}
%
\begin{eqnarray}
&&
I_{\downarrow\downarrow}=\frac{2e}{\hbar S}
\int d\epsilon d\omega\sum_{q,k1,k2}
\left|\sum_{k^\prime}
t_{SF}({\bf k}^\prime,{\bf k}_1)t_{SF}(-{\bf k}^\prime,{\bf k}_2)g_{SS|12}({\bf k^\prime})\right|^2
\left|\sum_p\frac{U_{qk2p}}{\xi_{p,\uparrow}-\xi_{k2,\downarrow}+\omega}\right|^2 
\times\label{Idd3}\\
&&
\times
\delta(\omega-\omega_q)
\delta(\epsilon+eV-\xi_{k1,\downarrow})
\delta(2eV-\xi_{k1,\downarrow}-\xi_{k2,\downarrow}+\omega)
\left\{
n(-\xi_{k1,\downarrow})n(-\xi_{k2,\downarrow})N(\omega)
-
n(\xi_{k1,\downarrow})n(\xi_{k2,\downarrow})[1+N(\omega)]
\right\}.
\nonumber
\end{eqnarray}
%
                   {\tiny
                   \noindent\begin{tabular}[t]{c|c}
                         \cline{2-2}  \\
                    \parbox{0.493\hsize}{~} &  \parbox{0.493\hsize}{~} \end{tabular}}\vspace*{-3.5ex}
\begin{multicols}{2}\noindent
The first term in Eq.~(\ref{ISub}) is the usual Andreev current $I_{\uparrow\downarrow}$ which,
as can be seen from Eq.~(\ref{Iud3}), results in either the gain or loss in the ferromagnet of two 
carriers belonging to different spin bands.
Their energies $\xi_{k1,\uparrow}$ and $\xi_{k2,\downarrow}$ measured with respect to 
the chemical potential on the F side $E_F$ obey a conservation law of the form
$\xi_{k1,\uparrow}+\xi_{k2,\downarrow}=2eV$ where $2eV$ is the energy of a Cooper pair with respect to $E_F$.
The current $I_{\uparrow\downarrow}$ is non-zero owing to the imbalance between the gain of the two electrons,  
$n(-\xi_{k1,\uparrow})n(-\xi_{k2,\downarrow})=[1-n(\xi_{k1,\uparrow})][1-n(\xi_{k2,\downarrow})]$, and their loss, 
$n(\xi_{k1,\uparrow})n(\xi_{k2,\downarrow})$.
In equilibrium ($V=0$) energy conservation ensures vanishing $I_{\uparrow\downarrow}$.
The factor
$|\sum_{k^\prime} t_{SF}({\bf k}^\prime,{\bf k}_1)t_{SF}(-{\bf k}^\prime,{\bf k}_2)g_{SS|12}({\bf k^\prime})|^2$
describes the two-particle tunneling process resulting in the creation or annihilation of a spinless Cooper pair 
with an anomalous average $g_{SS|12}({\bf k^\prime})$ on the S side. Using the selection rules 
expressed by the delta-functions in Eq.~(\ref{Iud3}), one can find that at zero temperature, $T=0$, the Andreev current 
(\ref{Iud3}) assumes the following form

\end{multicols}\vspace*{-3.5ex}{\tiny
                   \noindent\begin{tabular}[t]{c|}
                   \parbox{0.493\hsize}{~} \\ \hline \end{tabular}}

\begin{eqnarray}
I_{\uparrow\downarrow}=\frac{8\pi e}{h}\int_{-eV}^{eV} d\epsilon
\left|\sum_{k^\prime}
t_{SF}({\bf k}^\prime,{\bf k}_1)t_{SF}(-{\bf k}^\prime,{\bf k}_2)g_{SS|12}({\bf k^\prime},\epsilon)\right|^2
\delta(-\epsilon+eV-\xi_{k1,\uparrow})
\delta(2eV-\xi_{k1,\uparrow}-\xi_{k2,\downarrow}).
\label{IudT0}
\end{eqnarray}
%
                   {\tiny
                   \noindent\begin{tabular}[t]{c|c}
                         \cline{2-2}  \\
                    \parbox{0.493\hsize}{~} &  \parbox{0.493\hsize}{~} \end{tabular}}\vspace*{-3.5ex}
\begin{multicols}{2}\noindent

The magnon-assisted contributions to the subgap current (\ref{ISub}) are given by Eqs.~(\ref{Iuu3}) and 
(\ref{Idd3}). Let us analyze first the contribution $I_{\uparrow\uparrow}$ from the majority carriers.
In Eq.~(\ref{Iuu3}) the product of the occupation numbers 
$n(-\xi_{k1,\uparrow})n(-\xi_{k2,\uparrow})][1+N(\omega)]$ represents the gain of two spin up electrons and a 
magnon in the F side. This process can be viewed as the injection of two electrons with energies $\xi_{k1,\uparrow}$ 
and $\xi_{p,\downarrow}$ into the ferromagnet where the spin down electron has a finite lifetime because of the intra-atomic 
exchange interaction with the localized moments. Then the spin down electron emits a magnon of energy $\omega$ 
and occupies an empty state $\xi_{k2,\uparrow}$ in the spin up conduction band. Note that the energy in the denominator of 
Eq.~(\ref{Iuu3}), $\xi_{p,\downarrow}-\xi_{k2,\uparrow}-\omega$, is related to the inverse 
lifetime of the spin down electron in the virtual state.
The reverse kinetic process consists of the loss of two spin up electrons and a magnon from the F region:
$n(\xi_{k1,\uparrow})n(\xi_{k2,\uparrow})N(\omega)$. It is possible only at finite temperatures because one of the 
spin up carriers has to absorb a magnon in order to form the necessary intermediate spin down state before 
tunneling into the superconductor. Both processes obey an energy conservation law of the form 
$2eV=\xi_{k1,\uparrow}+\xi_{k2,\uparrow}+\omega$. Thus, if the energy $2eV$ of a Cooper pair is equal to zero 
(at $V=0$), the direct and reverse processes compensate each other. 

At zero temperature, where the majority current $I_{\downarrow\downarrow}$ (\ref{Iuu3}) is determined by the 
gain of electrons in the F side, the integration over the electron energy $\epsilon$ is reduced to the interval 
$\omega-eV\leq\epsilon\leq eV$. As a consequence, only magnons with energies less than the energy of a 
Cooper pair, i.e. $0<\omega<2eV$, contribute to the current. Furthermore, the last inequality shows that 
the majority current is non-zero only at positive voltages. The minimal value of $V$ is related to the gap 
energy $\omega_0$ in the magnon spectrum: $\omega_0/2e\leq V$.
On the other hand, the inequalities $\omega\leq 2eV\leq 2\Delta$ imply that the magnons 
of interest are mainly long-wavelength ones because the superconducting gap energy $\Delta$ is usually much less 
than the magnon Debye energy. In other words, in Eq. (\ref{Hem}) for the matrix element $U_{qk2p}$ of the 
electron-magnon interaction the magnon wave function varies much more slowly than the electron ones. For this 
reason, $U_{qk2p}$ falls rapidly as the difference $|{\bf p}-{\bf k}_2|$ in the electron momenta increases, which 
means that both the virtual (spin down) and the final (spin up) electron states are close to the majority Fermi surface. 
It allows us to neglect the change in the electron momentum in the denominator of Eq.~(\ref{Iuu3}) and calculate the 
sum over the momentum of the virtual state ${\bf p}$ as follows

\begin{eqnarray}
\sum_p\frac{U_{qk2p}}{\xi_{p,\downarrow}-\xi_{k2,\uparrow}-\omega} &\approx&
\frac{1}{\Gamma\gamma({\bf k}_2)}\sum_p U_{qk2p} =
\label{integral}\\
&=& \frac{u({\bf k}_2)}{\gamma({\bf k}_2)}L^{1/2}_F\chi^*_{q_z}(0).
\nonumber
\end{eqnarray}
In order to obtain the last equation we have used formula (\ref{Hem}) for $U_{qk2p}$ with wave functions 
for electrons in a large volume with a hard wall at $z=0$, i.e. 
$\Phi_k({\bf r})=2^{1/2}(AL_F)^{-1/2}\exp(i{\bf k}_{||}{\bf r}_{||})\sin(k_z z)$, and magnon wave functions of the 
form $\chi_q({\bf r})=A^{-1/2}\exp(i{\bf q}_{||}{\bf r}_{||})\chi_{q_z}(z)$, where $\chi_{q_z}(z)$ corresponds to
motion in the direction perpendicular to the contact plane. It can be shown that the integration over 
${\bf p}$ in Eq.~(\ref{integral}) gives rise to a $\delta$-shaped factor which in the limit $(k_z/q_z)\to\infty$ 
transforms into $\delta(z)$. That is why the effectiveness of the electron-magnon interaction depends on the 
magnitude of the magnon wave function near the interface (Eq.~(\ref{integral})). According to (\ref{integral}), 
the contribution of Eq.~(\ref{Iuu3}) can be represented for $T\to 0$ as

\end{multicols}\vspace*{-3.5ex}{\tiny
                   \noindent\begin{tabular}[t]{c|}
                   \parbox{0.493\hsize}{~} \\ \hline \end{tabular}}

\begin{eqnarray}
&&
I_{\uparrow\uparrow}=\frac{4\pi e}{hS}\Theta(eV-\omega_0)
\int_0^{2eV}d\omega \Omega(\omega)\int_{\omega -eV}^{eV}d\epsilon\sum_{k1,k2}
\frac{|u({\bf k}_2)|^2}{\gamma^2({\bf k}_2)}
\left|\sum_{k^\prime}
t_{SF}({\bf k}^\prime,{\bf k}_1)t_{SF}(-{\bf k}^\prime,{\bf k}_2)g_{SS|12}({\bf k^\prime})\right|^2
\times
\label{IuuT0}\\
&&
\qquad \qquad \qquad \qquad \qquad \qquad \qquad \qquad \qquad \qquad \qquad \qquad \times \, 
\delta(-\epsilon+eV-\xi_{k1,\uparrow})
\delta(2eV-\xi_{k1,\uparrow}-\xi_{k2,\uparrow}-\omega),
\nonumber
\end{eqnarray}
%
                   {\tiny
                   \noindent\begin{tabular}[t]{c|c}
                         \cline{2-2}  \\
                    \parbox{0.493\hsize}{~} &  \parbox{0.493\hsize}{~} \end{tabular}}\vspace*{-3.5ex}
\begin{multicols}{2}\noindent
where $\Omega(\omega)$ denotes the magnon local density of states at the interface

\begin{equation}
\Omega(\omega)=L_F\sum_q |\chi_{q_z}(0)|^2\delta(\omega-\omega_q).
\label{MDOS}
\end{equation}
Since the magnon wave function $\chi_{q_z}(z)$ is normalized on the length $L_F$ of the ferromagnet, 
the quantity $L_F|\chi_{q_z}(0)|^2$ in Eq. (\ref{MDOS}) is the dimensionless magnon amplitude at the interface
which does not depend on $L_F$.

The other magnon-assisted contribution $I_{\downarrow\downarrow}$ (\ref{Idd3}) is formed by minority electron states.
The kinetic processes giving rise to $I_{\downarrow\downarrow}$ consist of the gain or loss in the F region
of two spin down electrons. In the direct process (the gain) Andreev 
reflection results in the injection of both a spin up and a spin down electron with energies $\xi_{p,\uparrow}$ and 
$\xi_{k1,\downarrow}$ respectively. Then the former absorbs a magnon of energy $\omega$ and occupies a state 
$\xi_{k2,\downarrow}$ in the minority conduction band. This takes place only at finite temperatures. In the reverse 
process (the loss) one of the minority electrons emits a magnon and forms an intermediate spin up state with energy 
$\xi_{p,\uparrow}$. This allows it to tunnel into the superconductor along with another spin down carrier having
energy $\xi_{k1,\downarrow}$. Both processes obey the energy conservation law of the form 
$2eV=\xi_{k1,\downarrow}+\xi_{k2,\downarrow}-\omega$.
It can be shown in an analogous way as for $I_{\uparrow\uparrow}$ (\ref{IuuT0}) that only interactions with
long-wavelength magnons are of importance, which take place in the vicinity of the minority Fermi surface.
We note that, in contrast to the direct process, the reverse one is possible at zero temperature.
That is why the $T\to 0$ asymptotic form for the current 
(\ref{Idd3}) is determined by the `loss' term in Eq.~(\ref{Idd3}) and, therefore, it is non-zero only at negative 
voltages $V\leq -\omega_0/2e$:

\end{multicols}\vspace*{-3.5ex}{\tiny
                   \noindent\begin{tabular}[t]{c|}
                   \parbox{0.493\hsize}{~} \\ \hline \end{tabular}}

\begin{eqnarray}
&&
I_{\downarrow\downarrow} = -\frac{4\pi e}{hS}\Theta(-eV-\omega_0)
\int_0^{-2eV}d\omega \Omega(\omega)\int_{\omega +eV}^{-eV}d\epsilon\sum_{k1,k2}
\frac{|u({\bf k}_2)|^2}{\gamma^2({\bf k}_2)}
\left|\sum_{k^\prime}
t_{SF}({\bf k}^\prime,{\bf k}_1)t_{SF}(-{\bf k}^\prime,{\bf k}_2)g_{SS|12}({\bf k^\prime})\right|^2
\times
\label{IddT0} \\
&&
\qquad \qquad \qquad \qquad \qquad \qquad \qquad \qquad \qquad \qquad \qquad \qquad \qquad \times \,
\delta(\epsilon+eV-\xi_{k1,\downarrow})
\delta(2eV-\xi_{k1,\downarrow}-\xi_{k2,\downarrow}+\omega).
\nonumber
\end{eqnarray}
%
                   {\tiny
                   \noindent\begin{tabular}[t]{c|c}
                         \cline{2-2}  \\
                    \parbox{0.493\hsize}{~} &  \parbox{0.493\hsize}{~} \end{tabular}}\vspace*{-3.5ex}
\begin{multicols}{2}\noindent
In Eqs.~(\ref{IudT0}), (\ref{IuuT0}) and (\ref{IddT0}) the anomalous Green function
$g_{SS|12}({\bf k^\prime},\epsilon)$ of the
superconductor is given by the well-known formula\cite{AGD} (see also Eq.~(\ref{homoS}))
which can be rewritten as follows

\begin{equation}
g_{SS|12}({\bf k^\prime},\epsilon)=
\frac{\Delta}{(\Delta^2-\epsilon^2)^{1/2}}\frac{(\Delta^2-\epsilon^2)^{1/2}}
{(\epsilon_S({\bf k}^\prime)-E_S)^2+\Delta^2-\epsilon^2}. 
\label{g12}
\end{equation}
Here the second multiplier represents the Lorentz curve as a function of $\epsilon_S({\bf k}^\prime)-E_S$ for 
subgap excitations with $\epsilon<\Delta$. In the Andreev approximation, where $\Delta\ll E_S$, it takes the 
delta-shaped form  

\begin{equation}
g_{SS|12}({\bf k}^\prime,\epsilon)=
\frac{\pi\Delta}{(\Delta^2-\epsilon^2)^{1/2}}\delta
(E_S-\epsilon_S({\bf k}^\prime)),
\label{g12A}
\end{equation}
that allows us to decouple the integrations over the energy $\epsilon$ and the electron wave vector ${\bf k}^\prime$ 
in Eqs.~(\ref{IudT0}), (\ref{IuuT0}) and (\ref{IddT0}). This means that the $I(V)$ characteristics of the contact turn out 
to be independent of the detailed structure of the interface.
Also, taking into account that $|\epsilon|\leq |eV|\ll \Gamma,E_S,E_F$, we have for the usual Andreev current

\begin{eqnarray}
I_{\uparrow\downarrow} &=& \frac{4e}{h}\Delta \, {\rm arctanh}\frac{eV}{\Delta}
\times
\label{IA}\\
&& \quad
\times
\left(
\frac{A}{h^2}
\right)^2
\int\limits_{\Pi_{\uparrow}}d{\bf p}_{||1}
\int\limits_{\Pi_{\downarrow}}d{\bf p}_{||2}\,
R_A({\bf p}_{||1},{\bf p}_{||2}),
\nonumber
\end{eqnarray}
\begin{equation}
\Pi_{\alpha}=\int d{\bf p}_{||}
\Theta(E_F-\epsilon_{\alpha}({\bf p}_{||},p_z=p_{\alpha,max})).
\label{PF}
\end{equation}
Here $\Pi_{\uparrow}$ and $\Pi_{\downarrow}$ are the areas of the maximal cross sections of the majority and minority 
Fermi surfaces in the plane parallel to the interface. The values $p_{\uparrow,max}$ and $p_{\downarrow,max}$ of the 
perpendicular momentum correspond to the positions of the maximal cross sections in the ${\bf p}$-space. The matrix 
$R_A({\bf p}_{||1},{\bf p}_{||2})$ in (\ref{IA}) describes Andreev reflection of a spin up channel specified by 
the value of parallel momentum ${\bf p}_{||1}$ into a spin down channel with parallel momentum ${\bf p}_{||2}$ 
or vice versa. The integrals in Eq.~(\ref{IA}) take into account all possible transitions between the spin up and 
spin down modes in the contact. The form of the reflection matrix depends on the quality of the interface and is 
given by

\begin{eqnarray}
R_A({\bf p}_{||1},{\bf p}_{||2}) &=& 
\left(\frac{A}{h^2}\right)^2
|
\int\limits_{\Pi_S}d{\bf p}^\prime_{||}\,
t({\bf p}^\prime_{||},{\bf p}_{||1})t(-{\bf p}^\prime_{||},{\bf p}_{||2})
|^2, \nonumber \\
\label{RA}
\end{eqnarray}
\begin{equation}
\Pi_{S}=\int d{\bf p}_{||}
\Theta(E_S-\epsilon_S({\bf p}_{||},p_z=p_{S,max})) ,
\label{PS}
\end{equation}
where $\Pi_S$ is the area of the maximal cross section of the Fermi surface in the superconductor.
The expression for the reflection matrix (\ref{RA}) can be derived by integrating over the perpendicular
momenta in Eq.~(\ref{IudT0}) with the use of the delta-functions and the formula (\ref{t}) for the
tunneling matrix elements. 

The inelastic contributions to the subgap $I(V)$ characteristics are given within the Andreev approximation by 

\end{multicols}\vspace*{-3.5ex}{\tiny
                   \noindent\begin{tabular}[t]{c|}
                   \parbox{0.493\hsize}{~} \\ \hline \end{tabular}}

\begin{eqnarray}
I_{\uparrow\uparrow}=\frac{e}{hS}
\left(\frac{A}{h^2}\right)^2
& &
\int\limits_{\Pi_{\uparrow}}d{\bf p}_{||1}
\int\limits_{\Pi_{\uparrow}}d{\bf p}_{||2}\,
R_{in}({\bf p}_{||1},{\bf p}_{||2})
\times
\label{Imaj}\\
&\times&
\Delta\Theta(eV-\omega_0)
\int_{0}^{2eV}\Omega(\omega)d\omega
\left[
{\rm arctanh}\frac{eV}{\Delta}+{\rm arctanh}\frac{eV-\omega}{\Delta}
\right],
\nonumber
\end{eqnarray}
%

\begin{eqnarray}
I_{\downarrow\downarrow}=\frac{e}{hS}
\left(\frac{A}{h^2}\right)^2
& &
\int\limits_{\Pi_{\downarrow}}d{\bf p}_{||1}
\int\limits_{\Pi_{\downarrow}}d{\bf p}_{||2}\,
R_{in}({\bf p}_{||1},{\bf p}_{||2})
\times
\label{Imin}\\
&\times&
\Delta\Theta(-eV-\omega_0)
\int_{0}^{-2eV}\Omega(\omega)d\omega
\left[
{\rm arctanh}\frac{eV}{\Delta}+{\rm arctanh}\frac{eV+\omega}{\Delta}
\right],
\nonumber
\end{eqnarray}
%
\begin{eqnarray}
R_{in}({\bf p}_{||1},{\bf p}_{||2})=\left(\frac{A}{h^2}\right)^2
\frac{|u({\bf p}_{||2})|^2}{\gamma^2({\bf p}_{||2})}
|
\int\limits_{\Pi_S}d{\bf p}^\prime_{||}\,
t({\bf p}^\prime_{||},{\bf p}_{||1})t(-{\bf p}^\prime_{||},{\bf p}_{||2})
|^2.
\label{Rin}
\end{eqnarray}
%
                   {\tiny
                   \noindent\begin{tabular}[t]{c|c}
                         \cline{2-2}  \\
                    \parbox{0.493\hsize}{~} &  \parbox{0.493\hsize}{~} \end{tabular}}\vspace*{-3.5ex}
\begin{multicols}{2}\noindent
Since a spin-flip process associated with magnon emission compensates the change in spin resulting from Andreev 
reflection, the matrix $R_{in}({\bf p}_{||1},{\bf p}_{||2})$ in Eqs.~(\ref{Imaj}) and (\ref{Imin}) 
for the magnon-assisted currents describes transitions between electron channels of the same spin. Note that 
$R_{in}({\bf p}_{||1},{\bf p}_{||2})$ is proportional to the factor $|u({\bf p}_{||2})|^2/\gamma^2({\bf p}_{||2})$ 
that takes into account an anisotropy of the intra-atomic exchange interaction near the Fermi surface of the majority 
and minority electrons. 

\subsection{Uniformly transparent interface}

Let us analyze now the properties of the matrices (\ref{RA}) and (\ref{Rin}) for a plane, wide contact with a clean
barrier where electron momenta parallel to the interface are conserved upon tunneling. In this case the factor 
$t({\bf p}^\prime_{||},{\bf p}_{||})$ in Eqs.~(\ref{RA}) and (\ref{Rin}) is proportional to the matrix element of 
a transition between two plane waves normalized by the contact area $A$:

\begin{equation}
t({\bf p}^\prime_{||},{\bf p}_{||})=t\frac{h^2\delta({\bf p}^\prime_{||}-{\bf p}_{||})}{A},
\label{tclean}
\end{equation}
where $|t|^2$ represents the tunneling probability per channel.
We assume that $|t|^2$ is much less than unity and is related to the 
barrier width which is, in turn, considered to be the same for all points along the contact.
For a contact with large, but finite area the peak of the function
$\delta({\bf p}^\prime_{||}-{\bf p}_{||})$ at ${\bf p}^\prime_{||}={\bf p}_{||}$ is as high as the ratio $A/h^2$. 
For this reason the factor $h^2\delta({\bf p}^\prime_{||}-{\bf p}_{||})/A$ in Eq.~(\ref{tclean}) should be 
treated as a Kronecker delta.
Assuming for simplicity that the maximal cross section $\Pi_S$ of the Fermi surface in the superconductor to be
the largest one: $\Pi_S>\Pi_{\uparrow}>\Pi_{\downarrow}$, we obtain matrices (\ref{RA}) and (\ref{Rin}) in the form

\begin{eqnarray}
&&
R_A({\bf p}_{||1},{\bf p}_{||2})=|t|^4 \frac{h^2\delta({\bf p}_{||1}+{\bf p}_{||2})}{A},
\label{Runi}\\
&&
R_{in}({\bf p}_{||1},{\bf p}_{||2})=R_A({\bf p}_{||1},{\bf p}_{||2})
\frac{|u({\bf p}_{||2})|^2}{\gamma^2({\bf p}_{||2})}.
\nonumber
\end{eqnarray}
As it should be for a homogeneous interface, the relationship between the parallel momenta of the incident and Andreev 
reflected particles on the F side is the same as for electrons forming a Cooper pair on the S side: 
${\bf p}_{||1}+{\bf p}_{||2}=0$.

\subsection{Randomly transparent interface}

Now we introduce a model for describing a strongly nonuniform interface which is transparent only in a finite 
number of points randomly distributed over an area $A$ with density $n_h$. The typical distance $n_h^{-1/2}$ between
the transparent points is assumed to be much 
larger than the Fermi wavelengths in the superconductor $\lambda_S$ and the ferromagnet $\lambda_F$:

\begin{equation}
\lambda_{S,F}n_h^{1/2}\ll 1.
\label{nh}
\end{equation}
Each transparent point contact can be treated as a defect which causes electron scattering in the plane parallel to
the interface. 
Under condition (\ref{nh}) we can describe the system of such defects by a sum of short-range potentials 
$a\sum_{j}t_j\delta({\bf r}-{\bf r}_j)$, where ${\bf r}_j$ is the position vector of the $j$th contact in the plane,
$a\sim\lambda_{S,F}^2$ is its area, and $t_j$ introduces the magnitude of the potential at ${\bf r}={\bf r}_j$.
The transfer matrix element $t({\bf p}^\prime_{||},{\bf p}_{||})$ (see Eqs.~(\ref{RA}) and (\ref{Rin})), which
accounts for the change in the parallel momentum of an electron upon tunneling, should be identified as the 
matrix element of the total scattering potential calculated with the use of plane waves normalized on $A$:

\begin{eqnarray}
t({\bf p}^\prime_{||},{\bf p}_{||}) &=& \int_A\frac{ad{\bf r}}{A}\sum_{j}t_j\delta({\bf r}-{\bf r}_j)
\exp[i\hbar^{-1}({\bf p}_{||}-{\bf p}^\prime_{||}){\bf r}]
\nonumber\\
&=&\frac{a}{A}
\sum_{j}t_j
\exp[i\hbar^{-1}({\bf p}_{||}-{\bf p}^\prime_{||}){\bf r}_j].
\label{tdis}
\end{eqnarray}
Here the constants $t_j$ are chosen to be dimensionless. 

Since the reflection matrices (\ref{RA}) and (\ref{Rin}) now become random functions of ${\bf r}_j$, we have to average them 
over the position of each defect in the same spirit as in the problem of electron-impurity scattering in bulk 
metals (see, e.g., Ref.~\onlinecite{AGD} and Refs therein).
The product of matrix elements to be averaged (see Eqs.~(\ref{RA}) and (\ref{Rin})) is  

\begin{eqnarray}
&&
\overline{
t({\bf p}^\prime_{||1},{\bf p}_{||1})t(-{\bf p}^\prime_{||1},{\bf p}_{||2})
t^*({\bf p}^\prime_{||2},{\bf p}_{||1})t^*(-{\bf p}^\prime_{||2},{\bf p}_{||2})
}
=
\nonumber\\
&&
\quad = \frac{a^4}{A^4}\sum_{j1,j2,j3,j4}t_{j1}t_{j2}t^*_{j3}t^*_{j4} \times
\nonumber\\
&&
\quad \quad \times \, \overline{
\exp i\hbar^{-1}\left[
({\bf p}_{||1}-{\bf p}^\prime_{||1}){\bf r}_{j1}+({\bf p}_{||2}+{\bf p}^\prime_{||1}){\bf r}_{j2}-
\right.
}
\nonumber\\
&&
\quad \quad \overline{
\left.
-({\bf p}_{||1}-{\bf p}^\prime_{||2}){\bf r}_{j3}-({\bf p}_{||2}+{\bf p}^\prime_{||2}){\bf r}_{j4}
\right]
},
\label{aver}
\end{eqnarray}
where the bar represents averaging. The sum in the argument of the exponential function is the total phase shift caused by 
scattering at the contacts located at points ${\bf r}_{j1}, {\bf r}_{j2}, {\bf r}_{j3}$ and ${\bf r}_{j4}$. If these 
points correspond to the same defect, i.e.  ${\bf r}_{j1}={\bf r}_{j2}={\bf r}_{j3}={\bf r}_{j4}$, the total phase shift is 
equal to zero. As a result, the contribution from such terms is $(a/A)^4\sum_j|t_j|^4$. Next, one has to average the product 
of the phase factors in (\ref{aver}) divided into all possible pairs ${\bf r}_{j1}={\bf r}_{j2}, {\bf r}_{j3}={\bf r}_{j4}$, 
... It is easy to show that the contribution from these terms is small compared to the contribution mentioned above
by the parameter $(\lambda_{S,F})^2n_h\ll 1$.
Averaging of the four independent phase factors in Eq.~(\ref{aver}) gives an even smaller result.
Thus, the averaged reflection matrices (\ref{RA}) and (\ref{Rin}) are

\begin{eqnarray}
&&
\overline{
R_A({\bf p}_{||1},{\bf p}_{||2})}=\left(\frac{a}{A}\right)^2\left(\frac{a\Pi_S}{h^2}\right)^2\sum_j|t_j|^4,
\label{Rdis}\\
&&
\overline{
R_{in}({\bf p}_{||1},{\bf p}_{||2})}=\overline{R_A}
\frac{|u({\bf p}_{||2})|^2}{\gamma^2({\bf p}_{||2})}, \quad \overline{R_A}\ll 1.
\nonumber
\end{eqnarray}
The matrix $\overline{R_A}$ for the usual Andreev process is independent of the momenta and proportional to the 
squared number of transverse electron modes $a\Pi_S/h^2\sim 1$ in one point contact on the S side.
The factor $\sum_j|t_j|^4$ gives the total transparency of the interface summed over all the contacts.
Since we have treated tunneling as a perturbation, each tunneling probability $|t_j|^4$ must be much 
less than unity.
Note that the momentum dependence of the inelastic scattering matrix $\overline{R_{in}}$ 
is related only to the anisotropy of the electron-magnon interaction.

\section{Discussion of the magnon-assisted conductance of a ferromagnet-superconductor junction}
\label{Cond}

The band structure of a ferromagnetic electrode near the Fermi level may have a dramatic effect on the 
subgap conductance of an FS contact. Let us demonstrate this by considering the extreme example
of a junction involving a half-metallic ferromagnet.

In half-metallic ferromagnets the splitting $\Gamma$ between the majority and minority subbands is so strong 
that the bottom energy $\Gamma\gamma(0)/2$ of the minority subband (see Eq.~(\ref{Hm})) is higher than the Fermi 
level, i.e. $\Gamma\gamma(0)>2E_F$. For this reason only majority carriers are present 
at the Fermi level, and, therefore, usual Andreev reflection is forbidden by the spin current conservation law.
Indeed, according to Eq.~(\ref{IA}), the usual Andreev current vanishes as the number of the minority spin channels 
$A\Pi_{\downarrow}/h^2$ goes to zero. The same is true for $I_{\downarrow\downarrow}$ (\ref{Imin}) 
because this magnon-assisted contribution is determined by the minority states near the Fermi energy. 
Thus, the current in the half-metallic electrode is completely spin polarized 
(mediated by the spin up carriers) and its value $I$ is equal to the inelastic magnon-assisted contribution 
(\ref{Imaj}): $I=I_{\uparrow\uparrow}$. 

%
\begin{figure}[t]
\hspace{0.02\hsize}
\epsfxsize=0.85\hsize
\epsffile{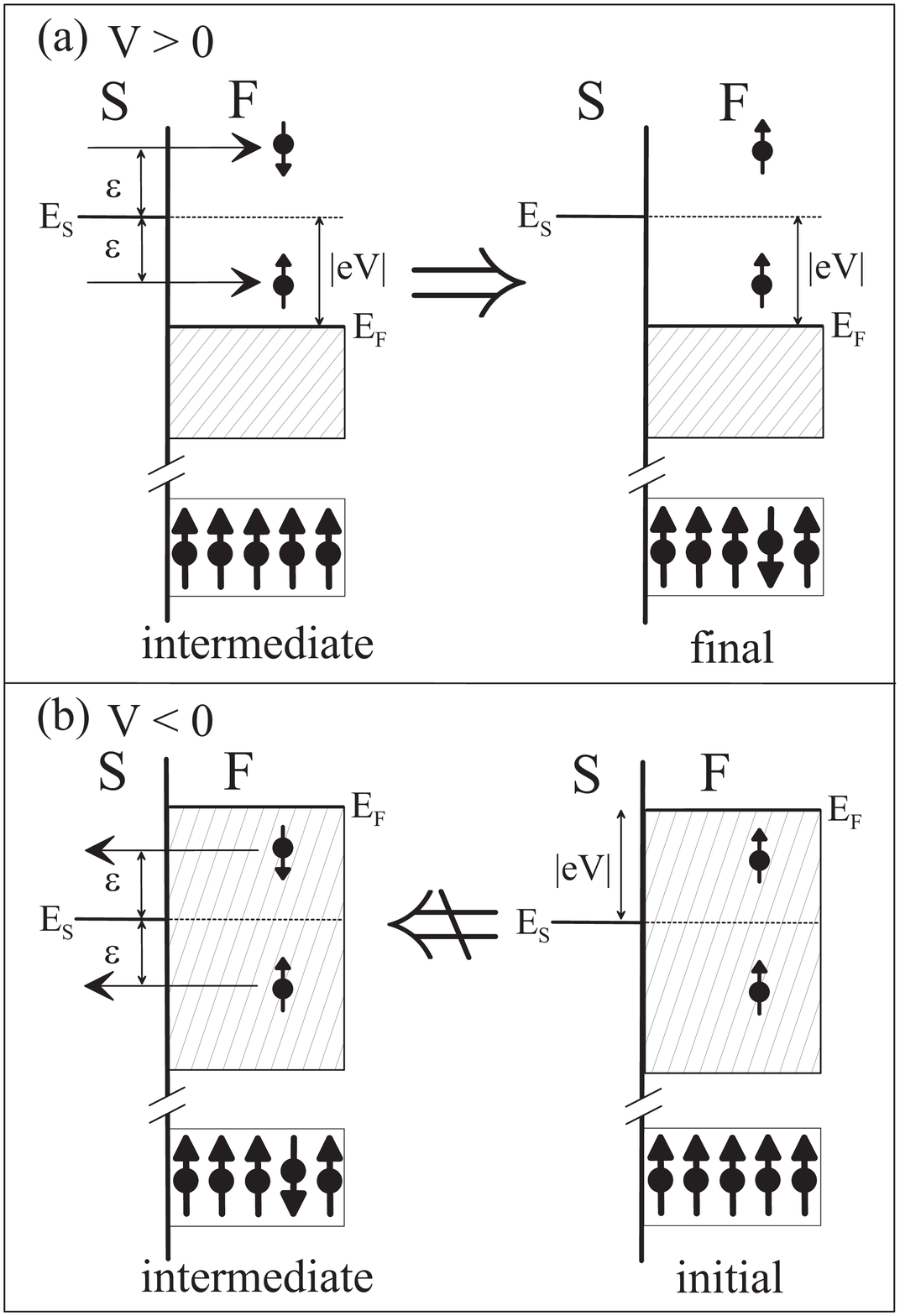}
\vspace{0.1cm}
\refstepcounter{figure}
\label{figure:3}

{\setlength{\baselineskip}{10pt} FIG.\ 3 - Tunneling process
between a superconducting electrode on the left hand side and a
half-metallic ferromagnet on the right for (a) $V>0$ and (b) $V<0$. For (a)
$V>0$ a down spin electron tunnelling into the ferromagnet may emit a magnon
and incorporate itself into the majority (up) conduction band. For (b) $V<0$ no
spin flip process is possible at $T=0$ because in the initial state (right)
there are no thermal magnons for an up spin electron to absorb.
}
\end{figure}
%
According to Eq.~(\ref{Imaj}), the $I(V)$ characteristics of the contact are asymmetric with respect to the bias voltage
$V$. The current is zero for negative bias and finite for $V>\omega_0/2e$. This asymmetry can be explained using the sketch
in Fig.~3 which shows the tunneling process between a superconducting electrode and a half-metallic ferromagnet 
for (a) $V>0$ and (b)  $V<0$. As discussed in the Introduction, for $V>0$, Fig.~3(a), Andreev reflection 
manifests itself as the injection of both a majority and a minority electron into the ferromagnet where they have 
energies $eV\pm\epsilon $ with respect to the Fermi level $E_{F}$. Since there are no spin down states near the Fermi 
level in a half-metallic ferromagnet, the state of the minority electron turns out to be virtual and the process of 
magnon emission allows this electron to occupy an empty state in the majority conduction band above $E_F$.

For $V<0$, Fig.3~(b), Andreev reflection would result in the injection of a singlet pair of electrons from the feromagnet 
into the superconductor. With respect to the Fermi level in the ferromagnet $E_F$ these energies are 
$-|eV|+\epsilon$ and $-|eV|-\epsilon$, respectively. A spin up electron cannot emit a 
magnon due to conservation of total spin in the exchange interaction so that the only possibility could be that a spin up 
electron in the ferromagnet would absorb a magnon to form an intermediate spin down state before tunneling into the 
superconductor. However there are no thermally excited magnons at $T=0$ in the initial state of the ferromagnet, so it 
is impossible for magnon-assisted Andreev reflection to contribute to current formation in negatively biased contacts. 

From Eqs.~(\ref{IuuT0}), (\ref{Runi}) and (\ref{Rdis}) we find that the inelastic differential conductance $G_{in}$ of a 
half-metal-superconductor contact is

\end{multicols}\vspace*{-3.5ex}{\tiny
                   \noindent\begin{tabular}[t]{c|}
                   \parbox{0.493\hsize}{~} \\ \hline \end{tabular}}

\begin{eqnarray}
G_{in}(V)=\frac{e^2}{hS}M(eV)
\int\limits_{\Pi_{\uparrow}}\frac{d{\bf p}_{||}}{\Pi_{\uparrow}}
\frac{|u({\bf p}_{||})|^2}{\gamma^2({\bf p}_{||})}\times
\left\{
\begin{array}{r@{\quad  \quad}l}
|t|^4(A\Pi_{\uparrow}/h^2)
& \mbox{ideal interface}\\
\sum_j|t_j|^4(a\Pi_S/h^2)^2(a\Pi_{\uparrow}/h^2)^2
& \mbox{disordered interface}
\end{array}
\right.,
\label{GinHFS}
\end{eqnarray}
                   {\tiny
                   \noindent\begin{tabular}[t]{c|c}
                         \cline{2-2}  \\
                    \parbox{0.493\hsize}{~} &  \parbox{0.493\hsize}{~} \end{tabular}}\vspace*{-3.5ex}
\begin{multicols}{2}\noindent
where the voltage-dependent factor $M(x)$ caused by the inelastic processes in the ferromagnet is defined in Eq.~(\ref{M}). 
Note that in the case of a disordered interface, the conductance (\ref{GinHFS}) is proportional to the product of four 
electron channels (two on either side of the contact) participating in transport through one point contact.  	
This means that, in the absence of the parallel momenta conservation law, two electron channels from one side of the 
junction are independently transformed into two channels on the other side.    

In the case of conventional ferromagnets the subgap current $I$ contains all three contributions (\ref{IudT0}), 
(\ref{IuuT0}) and (\ref{IddT0}), and, therefore, the differential conductance has both the elastic (Andreev) 
and inelastic parts: $G=G_A+G_{in}$.
For the case of a uniform interface, $G_A$ and $G_{in}$ can be obtained using Eqs.~(\ref{IudT0}), (\ref{IuuT0}), 
(\ref{IddT0}) and Eqs. (\ref{Runi}) for the reflection matrices $R_A$ and $R_{in}$.
The results are given by Eqs.~(\ref{GA}) -- (\ref{M}).
Calculating Andreev and magnon-assisted conductances for a disordered interface, we find that  

\begin{equation}
G_A(V)
=\frac{4e^2}{h}\sum_j|t_j|^4
\frac{a^4\Pi_S^2\Pi_{\uparrow}\Pi_{\downarrow}}{h^8(1-(eV/\Delta)^2)},
\label{GAdis}
\end{equation}
%
\end{multicols}\vspace*{-3.5ex}{\tiny
                   \noindent\begin{tabular}[t]{c|}
                   \parbox{0.493\hsize}{~} \\ \hline \end{tabular}}

\begin{eqnarray}
G_{in}(V)=\frac{e^2}{hS}\sum_j|t_j|^4\left(\frac{a\Pi_S}{h^2}\right)^2\left(\frac{a\Pi_{\uparrow}}{h^2}\right)^2
\left[
M(eV)
\int\limits_{\Pi_{\uparrow}}\frac{d{\bf p}_{||}}{\Pi_{\uparrow}}
\frac{|u({\bf p}_{||})|^2}{\gamma^2({\bf p}_{||})}
+
M(-eV)
\frac{\Pi_{\downarrow}}{\Pi_{\uparrow}}
\int\limits_{\Pi_{\downarrow}}\frac{d{\bf p}_{||}}{\Pi_{\uparrow}}
\frac{|u({\bf p}_{||})|^2}{\gamma^2({\bf p}_{||})}
\right].
\label{GinCFS}
\end{eqnarray}
                   {\tiny
                   \noindent\begin{tabular}[t]{c|c}
                         \cline{2-2}  \\
                    \parbox{0.493\hsize}{~} &  \parbox{0.493\hsize}{~} \end{tabular}}\vspace*{-3.5ex}
\begin{multicols}{2}\noindent
In contrast to the case of a half-metallic ferromagnet (see Eq.~(\ref{GinHFS})), in the general situation the inelastic conductance 
(\ref{GinCFS}) is non-zero at negative bias as well. As discussed in the Introduction and 
Section \ref{Sub}, this feature is explained by the presence of magnon-assisted transport of the minority carriers. 

We note that the electron and magnon `contributions' to the magnon-assisted conductance $G_{in}$ fall into 
separate multipliers (see, e.g., Eqs.~(\ref{GinHFS}) and (\ref{GinCFS})). On the one hand, this fact is due to the small 
momentum transfer of the electron-magnon interaction (see Eq.~(\ref{integral})). On the other hand, it is necessary that 
the superconducting gap energy is much less than Fermi energy, $\Delta\ll E_S$, (see Eq.~(\ref{g12A}) for the anomalous 
average under the Andreev approximation). Within these reasonable restrictions our approach allows us to calculate the tunnel 
conductance for arbitrary electron and magnon dispersion spectra and an arbitrary shaped potential 
barrier at the interface. Although the interfacial disorder does not influence the form of the $I(V)$ characteristics 
(see Eqs.~(\ref{IudT0}), (\ref{IuuT0}) and (\ref{IddT0})), it determines the dependence of the conductances
on the parameters of the Fermi surfaces in the superconductor and the ferromagnet (cf. Eqs.~(\ref{GA}), (\ref{Gin}) and 
Eqs.~(\ref{GAdis}), (\ref{GinCFS})). 

According to our analysis, the inelastic processes of magnon emission should manifest themselves in the asymmetric 
behaviour of the differential conductance $G(V)$, which is a direct consequence of the spin polarization of 
carriers in the ferromagnetic electrode. In order to observe the inelastic contribution to $G(V)$ it is necessary to 
eliminate the conventional Andreev (symmetric) part of the conductance (\ref{GA}), (\ref{GAdis}).
Besides, the proposed theory demands
that the gap in the magnon spectrum must be smaller than the superconducting gap energy, $\omega_0<\Delta$.
The latter requirement might not be satisfied in the experiments of Refs.~\onlinecite{sou98,pet99,L+G99,Gir98,Upa98}.
That is why the results of these works can be treated neither in favour nor against the existence of the
magnon-assisted Andreev processes.
On the other hand, the requirement $\omega_0<\Delta$ can be met in soft half-metallic materials, e.g., manganites \cite{Sr}. 
In Refs.\onlinecite{vas97,wei99} these compounds were used in tunnel contacts with high-$T_c$ superconductors 
(materials with large $\Delta$).
The $(dI/dV)-V$ curves measured in Ref.~\onlinecite{wei99} clearly demonstrated asymmetry 
for $|V|\leq\Delta/e$ which is consistent with that shown by Eqs.~(\ref{Gin}) and (\ref{GinCFS}).
As pointed out in Ref.~\onlinecite{wei99}, the asymmetry of the $(dI/dV)-V$ curves might be attributed to the $d$-wave type 
pairing in the high-$T_c$ superconductor.
However $d$-wave simulations performed in Ref.~\onlinecite{wei99} gave good results for quite large voltages
$|V|\approx\Delta/e$, but did not explain the asymmetry in the range $|V|<\Delta/e$ where the 
magnon-assisted Andreev reflection may occur. 

%
The authors thank I.~L.~Aleiner, B.~L.~Altshuler, N.~R.~Cooper, D.~M.~Edwards,
C.~J.~Lambert and Yu.~V.~Nazarov for discussions.
This work was supported by EPSRC, NATO and the Royal Society.
%
\begin{appendix}

\section{Keldysh functions formalism for calculating subgap tunneling current}
\label{Form}

Following Ref.~\onlinecite{Cuevas}, we introduce the $2\times 2$ Nambu representation for the nonequilibrium (Keldysh) 
Green function in the system of units where $\hbar=1$:

\begin{eqnarray}
&&
i{\cal G}_{EE^\prime}^{j ,j^\prime}(t,t^\prime)=
\label{G}\\
&&
=\left(
\begin{array}{cc}
<T_K c_{E\uparrow}(t_j)c^\dag_{E^\prime\uparrow}(t^\prime_{j^\prime})>&
<T_K c_{E\uparrow}(t_j)c_{E^\prime\downarrow}(t^\prime_{j^\prime})>\\
<T_K c^\dag_{E\downarrow}(t_j)c^\dag_{E^\prime\uparrow}(t^\prime_{j^\prime})>&
<T_K c^\dag_{E\downarrow}(t_j)c_{E^\prime\downarrow}(t^\prime_{j^\prime})>
\end{array}
\right).
\nonumber
\end{eqnarray}
Here $T_K$ is an operator  ordering the times along the Keldysh  contour which consists of a ``positive''  branch 
going from  $-\infty$ to $+\infty$  and a ``negative'' one going from  $+\infty$ to  $-\infty$.  Any time of the 
negative  branch  is  considered  as  posterior  to  any  time  of  the  positive  branch.
The index $j (j^\prime)$ means ``$+$'' or ``$-$'' depending on which branch of the Keldysh contour the time argument 
$t_j (t^{\prime}_{j^\prime})$ belongs to.
The subscripts $E$ and $E^\prime$ label the electrodes (S or F) in which a particle is created or annihiliated.
Using these definitions, it is easy to check that, for instance, the Keldysh function
${\cal G}^{++}_{FF|11}   =-i<T_K   a_{\uparrow}(t)a^\dag_{\uparrow}(t^\prime)>$
contains chronological ordering of the operators and, thus, it is the usual causal Green function of the spin up
electrons in the ferromagnet.
We also note that according to Eq.~(\ref{G}) the functions
${\cal G}^{+-}_{EE^\prime|\alpha\alpha^\prime}(t,t^\prime)$  and  
${\cal G}^{-+}_{EE^\prime|\alpha\alpha^\prime}(t,t^\prime)$  are anti-hermitian, i.e.,

\begin{eqnarray}
&
{\cal G}^{+-}_{EE^\prime|\alpha\alpha^\prime}(t,t^\prime)^*=
-{\cal G}^{+-}_{E^\prime ,E|\alpha^\prime\alpha}(t^\prime ,t),
&\label{anti}\\
&
{\cal G}^{-+}_{EE^\prime|\alpha\alpha^\prime}(t,t^\prime)^*=
-{\cal G}^{-+}_{E^\prime ,E|\alpha^\prime\alpha}(t^\prime ,t).
&\nonumber
\end{eqnarray}
Integration over the Keldysh contour is reduced to the interval $-\infty$ to $+\infty$ as follows

\begin{equation}
\int_Kdt...=\int_{-\infty}^{+\infty}dt\sum_{j=+1,-1}j...
\label{IntKeld}
\end{equation}

Since we deal with a static  problem, it is  convenient to go over to the energetic representation for Green functions 
${\cal G}^{+-}_{FS|11}$ and ${\cal G}^{-+}_{FS|22}$ in (\ref{IKeld}). Then they can be calculated 
using conventional perturbative diagram technique (see, e.g., Ref.~\onlinecite{AGD}) where the perturbation Hamiltonians are 
given by Eqs.~(\ref{APert}) and (\ref{MAPert}). If we multiply diagrams for ${\cal G}^{+-}_{FS|11}$ and 
${\cal G}^{-+}_{FS|22}$ by $t_{SF}({\bf k}^\prime,{\bf k})$ and $t^*_{SF}({\bf k}^\prime,{\bf k})$ respectively 
and integrate over all momenta and energies, we will obtain diagrams in the form of closed loops which represent 
the perturbation series for the tunneling current $I$ (\ref{IKeld}). These diagrams are shown in Fig.~4.  

Solid lines in Fig.~4 represent unperturbed electron Green functions of the ferromagnet
$g^{jj^\prime}_{FF|11}$, $g^{jj^\prime}_{FF|22}$ and the superconductor $g^{jj^\prime}_{SS|12}$,  
$g^{jj^\prime}_{SS|21}$.  
Dashed lines are the unperturbed magnon functions $d^{jj^\prime}$.
The arrows indicate the order of the arguments (momenta and Keldysh indices $j, j^\prime=+,-$).
Each tunneling matrix element $t_{SF}$ couples Green functions of the different  electrodes. 
In other words, two solid lines connected by $t_{SF}$ represent an elementary hybrid block which gives one 
Nambu matrix element of FS or SF Green function to the first order in the tunneling constant. 
Since both the FS and SF functions are $2\times2$ matrices, there are 8 different hybrid blocks. Note that the 
diagrams shown in Fig.~4 contain only the off-diagonal in Nambu indices blocks which are relevant to the process of 
Andreev reflection. The electron energy $\epsilon$ is assumed to be measured from the Fermi level $E_S$ in the 
superconductor. Thus electron Green functions of the ferromagnet have $\epsilon+\alpha eV$ as the energy argument.

%
\begin{figure}[t]
\epsfxsize=0.95\hsize
\epsffile{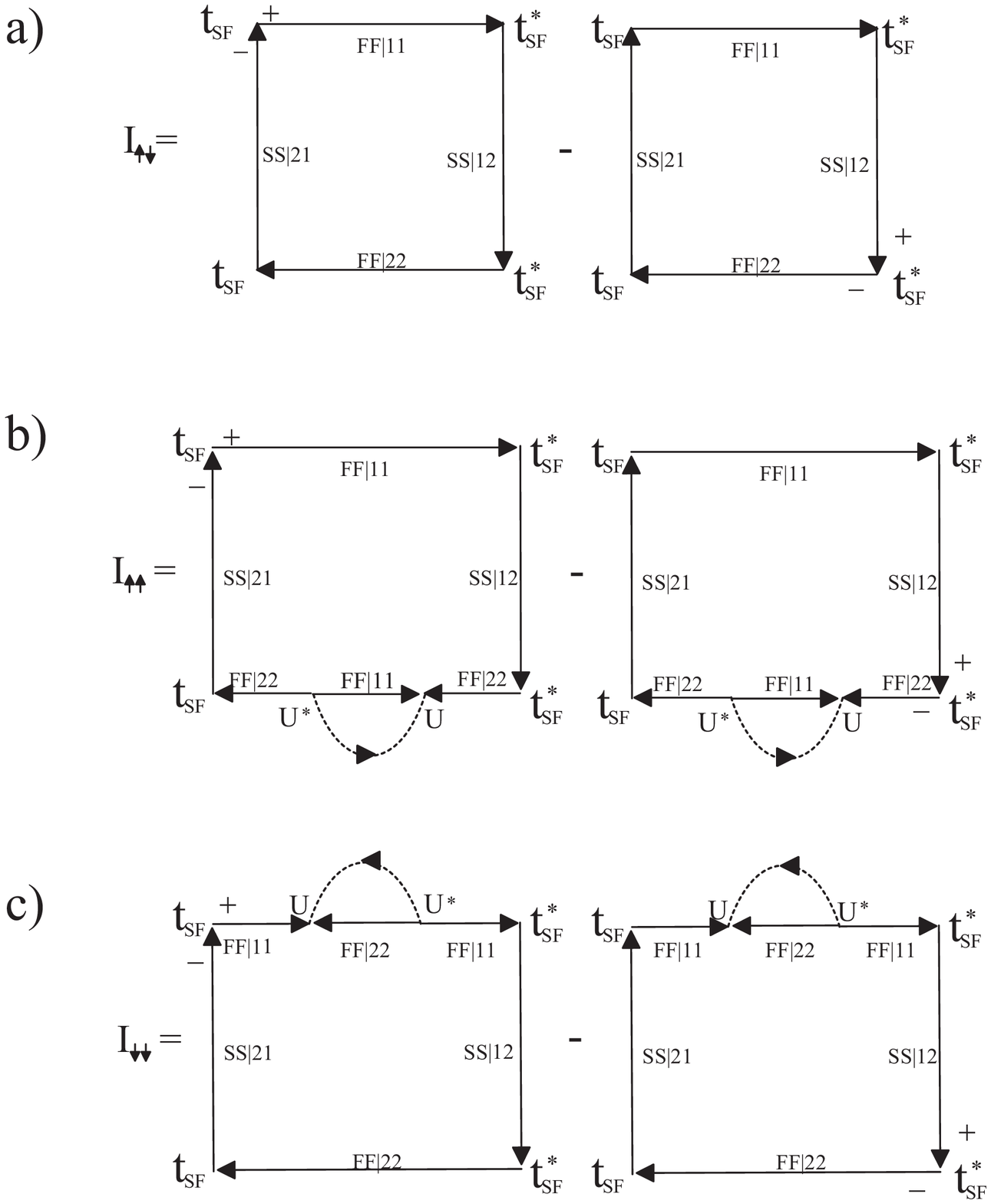}
\refstepcounter{figure}
\label{figure:4}

{\setlength{\baselineskip}{10pt} FIG.\ 4. - Diagrammatic representation of 
(a) Andreev current $I_{\uparrow \downarrow}$,
(b) magnon-assisted contribution from the majority 
carriers $I_{\uparrow \uparrow}$ and
(c) magnon-assisted contribution from the minority electrons
$I_{\downarrow \downarrow}$.
}
\end{figure}
%

The diagrams in Fig.~4(a) correspond to the usual Andreev process where the incoming electron from the ferromagnet
and the Andreev reflected one occupy the opposite spin bands. As we will see later, the sign ``$-$'' in Fig.~4 
ensures that only causal Green functions enter the expression for the current.
The diagrams in Figs.~4(b) and 4(c) describe Andreev reflection which is preceded or followed by spin-flip processes
caused by the emission or absorption of magnons.
These processes compensate the change in spin resulting from Andreev reflection.
That is why both the incoming and the outgoing electrons belong to the same spin band.
In the diagrams in Fig.~4(b) the initial and final electron states are both spin up ones.
They are described by Green functions with the subscripts $FF|11$. 
Green functions $FF|22$ correspond to intermediate (virtual) states of a spin down electron.
As pointed out in the previous sections, such a state can be formed as a result of Andreev reflection, then the 
spin down electron emits a magnon and incorporates itself into an empty state in the spin up conduction band. 
Alternatively, a virtual spin down state can be formed as a result of the absorption of a magnon by a spin up 
electron. Then this spin down state is Andreev reflected as a majority carrier. Since the formalism developed 
in this section does not require temperature to be zero, the diagrams in Fig.~4(b) 
take into account both of these processes which lead to the existence of `gain' and `loss' terms in 
the equation for the majority magnon-assisted current (see Eq.~(\ref{Iuu3}) in Section \ref{Sub}). 
In the diagrams in Fig.~4(c) both the incoming and outgoing electrons belong to the minority spin band. 
They are represented by Green functions with subscripts $FF|22$. 
Such an electron can either emit a magnon and form an intermediate spin up state, which then experiences Andreev reflection, 
or first be Andreev reflected as a majority carrier which then absorbs a magnon. These competing processes result in the 
`gain'-`loss' structure of the minority magnon-assisted current (see Eq.~(\ref{Idd3}) in Section \ref{Sub}).

Each Green function in Fig.~4 is expressed in terms of the retarded $g^R$, advanced  $g^A$ and Keldysh 
$g^K$ functions ($d^R$, $d^A$ and $d^K$ for magnons) by means of the relationship

\begin{equation}
g^{jj^\prime}=(g^K+j g^A+j^\prime g^R)/2,\qquad
j,j^\prime=+,-,
\label{KAR} 
\end{equation}
which together with (\ref{IntKeld}) allows us to formalize summations over the Keldysh indices. Indeed, 
at a given vertex labeled by the index $j$ the Keldysh sum has the structure $\sum_{j=+1,-1}j^{1+m}$, 
where the factor $j^m$ ($m$ is integer) comes from the product of Green functions (\ref{KAR}) connected 
by this vertex. It is clear that only those products of Green functions which give odd powers of $j$ 
`survive' after the summation: 

$$
\sum_{j=+1,-1}j^{1+m}=
\begin{array}{r@{\, \,}l}
2 & \mbox{for odd m}\\
0 & \mbox{for even m}
\end{array}.
$$
The Keldysh functions $g^K$ and $d^K$ are related to retarded and advanced functions as follows: 

\begin{eqnarray}
g^K_{FF|\alpha\alpha} &=& [g^R_{FF|\alpha\alpha}(\epsilon+\alpha eV)
-g^A_{FF|\alpha\alpha}(\epsilon+\alpha eV)]
\times \! \label{gKdK}\\
&& \qquad \times[1-2n(\epsilon+\alpha eV)],\nonumber\\
g^K_{SS|\alpha\alpha^\prime} &=& [g^R_{SS|\alpha\alpha^\prime}(\epsilon)-g^A_{SS|\alpha\alpha^\prime}(\epsilon)]
[1-2n(\epsilon)], \nonumber\\
d^K(\omega) &=& [d^R(\omega)-d^A(\omega)][1+2N(\omega)]
\nonumber
\end{eqnarray}
with $n$ and $N$ denoting the electron and magnon occupation numbers respectively and $\omega$ is the magnon energy. 

After some algebra the tunneling current $I$ (\ref{IKeld}) can be represented as the sum of three 
different contributions (see Eq.~(\ref{ISub})) given by

\begin{eqnarray}
&&
I_{\uparrow\downarrow}=eRe\int\frac{d\epsilon}{2\pi}\sum_{[i1..i4]}
{\rm Tr}\left\{
\right.
\label{Iud1}\\
&&
g^{i_1}_{FF|11}t^*_{SF}g^{i_2}_{SS|12}t^*_{SF}g^{i_3}_{FF|22}t_{SF}g^{i_4}_{SS|21}t_{SF}-
\nonumber\\
&&
\left.
-g^{i_1}_{FF|22}t_{SF}g^{i_2}_{SS|21}t_{SF}g^{i_3}_{FF|11}t^*_{SF}g^{i_4}_{SS|12}t^*_{SF}
\right\},
\nonumber
\end{eqnarray}
\begin{eqnarray}
&&
I_{\uparrow\uparrow}=eRe\int\frac{d\epsilon}{4\pi}\sum_{[i1..i6]}
{\rm Tr}\left\{
\right.
\label{Iuu1}\\
&&
g^{i_1}_{FF|11}t^*_{SF}g^{i_2}_{SS|12}t^*_{SF}g^{i_3}_{FF|22}\Sigma^{i_4}_{22}g^{i_5}_{FF|22}
t_{SF}g^{i_6}_{SS|21}t_{SF}-
\nonumber\\
&&
\left.
-g^{i_1}_{FF|22}\Sigma^{i_2}_{22}g^{i_3}_{FF|22}
t_{SF}g^{i_4}_{SS|21}t_{SF}g^{i_5}_{FF|11}t^*_{SF}g^{i_6}_{SS|12}t^*_{SF}
\right\},
\nonumber
\end{eqnarray}
\begin{eqnarray}
&&
I_{\downarrow\downarrow}=eRe\int\frac{d\epsilon}{4\pi}\sum_{[i1..i6]}
{\rm Tr}\left\{
\right.
\label{Idd1}\\
&&
g^{i_1}_{FF|11}\Sigma^{i_2}_{11}g^{i_3}_{FF|11}
t^*_{SF}g^{i_4}_{SS|12}t^*_{SF}g^{i_5}_{FF|22}t_{SF}g^{i_6}_{SS|21}t_{SF}-
\nonumber\\
&&
\left.
-g^{i_1}_{FF|22}t_{SF}g^{i_2}_{SS|21}t_{SF}g^{i_3}_{FF|11}\Sigma^{i_4}_{11}g^{i_5}_{FF|11}
t^*_{SF}g^{i_6}_{SS|12}t^*_{SF}
\right\},
\nonumber
\end{eqnarray}
Expression~(\ref{Iud1}) involves the sum over the set of indices 
$[i_1..i_4]=$ $KAAA$, $RKAA$, $RRKA$, $RRRK$, i.e. each term in this sum contains one Keldysh function multiplyed 
by the advanced functions on the right and/or by the retarded ones on the left. The analogous sums in Eqs.~(\ref{Iuu1}) 
and (\ref{Idd1}) have six terms of the structure of $KA..A$, $RKA..A$,.. and $R..RK$. Such terms have poles 
in both half-planes of the complex variable $\epsilon$ and, therefore, give nonvanishing contributions to the current.
Each Green function and the hopping element in the curly brackets in Eqs.~(\ref{Iud1})--(\ref{Idd1}) depends on two momentum 
variables which are considered to be matrix indices (discrete, in the general case) so that the products in the curly 
brackets are treated as matrix ones. The symbol ${\rm Tr}$ means the sum of all the diagonal elements of these products. 
By $\Sigma^R_{\alpha\alpha}$, $\Sigma^A_{\alpha\alpha}$ and $\Sigma^K_{\alpha\alpha}$ we denote the self-energies 
for the electron-magnon interaction:

\begin{eqnarray}
&&
\Sigma^{R(A)}_{11}=-\frac{i}{2S}\sum_{q_1,q_2,k_1,k_2} \!
U_{q1,k1,p1}U^*_{q2,k2,p2}
\int\frac{d\omega}{2\pi}
\times
\label{SAR11}\\
&&
\times
\left(
d^K({\bf q}_2,{\bf q}_1,\omega)g^{A(R)}_{FF|22}({\bf k}_2,{\bf k}_1, -\epsilon-\omega-eV)
+
\right.
\nonumber\\
&&
\left.
+d^{A(R)}({\bf q}_2,{\bf q}_1,\omega)g^K_{FF|22}({\bf k}_2,{\bf k}_1, -\epsilon-\omega-eV)
\right),
\nonumber
\end{eqnarray}
%
\begin{eqnarray}
&&
\Sigma^{R(A)}_{22}=-\frac{i}{2S}\sum_{q_1,q_2,k_1,k_2} \!
U_{q1,k1,p1}U^*_{q2,k2,p2}
\int\frac{d\omega}{2\pi}
\times
\label{SAR22}\\
&&
\times
\left(
d^K({\bf q}_2,{\bf q}_1,\omega)g^{A(R)}_{FF|11}({\bf k}_2,{\bf k}_1, -\epsilon-\omega+eV)
+
\right.
\nonumber\\
&&
\left.
+d^{A(R)}({\bf q}_2,{\bf q}_1,\omega)g^K_{FF|11}({\bf k}_2,{\bf k}_1, -\epsilon-\omega+eV)
\right),
\nonumber
\end{eqnarray}
%
\begin{eqnarray}
&&
\Sigma^K_{11}=-\frac{i}{2S}\sum_{q_1,q_2,k_1,k_2} \!
U_{q1,k1,p1}U^*_{q2,k2,p2}
\int\frac{d\omega}{2\pi}
\times
\label{SK11}\\
&&
\times
\left(
d^R({\bf q}_2,{\bf q}_1,\omega)g^R_{FF|22}({\bf k}_2,{\bf k}_1, -\epsilon-\omega-eV)
+
\right.
\nonumber\\
&&
+d^{A}({\bf q}_2,{\bf q}_1,\omega)g^A_{FF|22}({\bf k}_2,{\bf k}_1, -\epsilon-\omega-eV)
+
\nonumber\\
&&
\left.
+d^K({\bf q}_2,{\bf q}_1,\omega)
g^K_{FF|22}({\bf k}_2,{\bf k}_1, -\epsilon-\omega-eV)
\right).
\nonumber
\end{eqnarray}
%
\begin{eqnarray}
&&
\Sigma^K_{22}=-\frac{i}{2S}\sum_{q_1,q_2,k_1,k_2}
U_{q1,k1,p1}U^*_{q2,k2,p2}
\int\frac{d\omega}{2\pi}
\times
\label{SK22}\\
&&
\times
\left(
d^R({\bf q}_2,{\bf q}_1,\omega)g^R_{FF|11}({\bf k}_2,{\bf k}_1, -\epsilon-\omega+eV)
+
\right.
\nonumber\\
&&
+d^{A}({\bf q}_2,{\bf q}_1,\omega)g^A_{FF|11}({\bf k}_2,{\bf k}_1, -\epsilon-\omega+eV)
+
\nonumber\\
&&
\left.
+d^K({\bf q}_2,{\bf q}_1,\omega)
g^K_{FF|11}({\bf k}_2,{\bf k}_1, -\epsilon-\omega+eV)
\right).
\nonumber
\end{eqnarray}
%
\begin{eqnarray}
&&
\Sigma^K_{\alpha\alpha}(\epsilon+\alpha eV)=[\Sigma^R_{\alpha\alpha}(\epsilon+\alpha eV)-
\Sigma^A_{\alpha\alpha}(\epsilon+\alpha eV)]
\times
\nonumber\\
&&
\times
[1-2n(\epsilon+\alpha eV)].
\label{sigmaK}
\end{eqnarray}
In the subgap regime $|eV|<\Delta$ Eqs. (\ref{Iud1})--(\ref{Idd1}) can be simplified. Since there no 
quasiparticle states in a superconductor in the energy range of interest ($|\epsilon|<|eV|<\Delta$), we can 
set $g^{R}_{SS|12}=g^{A}_{SS|12}=g_{SS|12}$ and $g^{R}_{SS|21}=g^{A}_{SS|21}=g_{SS|21}$ 
in Eqs.~(\ref{Iud1})--(\ref{Idd1}). In line with Eq.~(\ref{gKdK}) all terms containing the Keldysh functions of 
the superconductor, $g^K_{SS|12}$ and $g^K_{SS|21}$, vanish: 

\begin{eqnarray}
&&
I_{\uparrow\downarrow}=eRe\int\frac{d\epsilon}{\pi}[n(\epsilon+eV)-n(\epsilon-eV)]\times
\label{Iud2}\\
&&
\times
{\rm Tr}\left\{
(g^A_{FF|11}-g^R_{FF|11})t^*_{SF}g_{SS|12}t^*_{SF}
\times
\right.
\nonumber\\
&&
\left.
\times
(g^A_{FF|22}-g^R_{FF|22})t_{SF}g_{SS|21}t_{SF}
\right\},
\nonumber
\end{eqnarray}
\begin{eqnarray}
&&
I_{\uparrow\uparrow}=eRe\int\frac{d\epsilon}{2\pi}[n(\epsilon+eV)-n(\epsilon-eV)]\times
\label{Iuu2}\\
&&
{\rm Tr}\left\{
(g^A_{FF|11}-g^R_{FF|11})t^*_{SF}g_{SS|12}t^*_{SF}\times
\right.
\nonumber\\
&&
\times\left.
(g^A_{FF|22}\Sigma^A_{22}g^A_{FF|22}-g^R_{FF|22}\Sigma^R_{22}g^R_{FF|22})t_{SF}g_{SS|21}t_{SF}
\right\},
\nonumber
\end{eqnarray}
\begin{eqnarray}
&&
I_{\downarrow\downarrow}=eRe\int\frac{d\epsilon}{2\pi}[n(\epsilon+eV)-n(\epsilon-eV)]\times
\label{Idd2}\\
&&
{\rm Tr}\left\{ \!
(g^A_{FF|11}\Sigma^A_{11}g^A_{FF|11}- \! g^R_{FF|11}\Sigma^R_{11}g^R_{FF|11})t^*_{SF}g_{SS|12}t^*_{SF}\times
\right.
\nonumber\\
&&
\times\left.
(g^A_{FF|22}-g^R_{FF|22})t_{SF}g_{SS|21}t_{SF}
\right\}.
\nonumber
\end{eqnarray}
Further simplifications can be made in the case of the homogeneous electrodes which are described 
by Green functions of the form (see, e.g., Ref.~\onlinecite{AGD})

\begin{eqnarray}
g^{R(A)}_{FF|\alpha\alpha}({\bf k}_1,{\bf k}_2,\epsilon) &=& \frac{\delta_{{\bf k}_1,{\bf k}_2}}
{[\epsilon+\alpha(eV-\xi_{k1,\alpha})\pm i0]} ,
\label{homoF}\\
d^{R(A)}({\bf q}_1,{\bf q}_2,\omega) &=& \frac{\delta_{{\bf q}_1,{\bf q}_2}}
{[\omega-\omega_{q1}\pm i0]} ,
\label{homoM}\\
g_{SS|12}({\bf k}^\prime_1,{\bf k}^\prime_2) &=& g_{SS|12}({\bf k}^\prime_1)\delta_{-{\bf k}^\prime_1,{\bf k}^\prime_2},
\nonumber\\
g_{SS|21}({\bf k}^\prime_1,{\bf k}^\prime_2) &=& g_{SS|21}({\bf k}^\prime_2)\delta_{{\bf k}^\prime_1,-{\bf k}^\prime_2},
\nonumber\\
g_{SS|12}({\bf k}^\prime,\epsilon) &=& g_{SS|21}({\bf k}^\prime,\epsilon) =
\nonumber\\
&=& \frac{\Delta}{[(\epsilon_S({\bf k}^\prime)-E_S)^2+\Delta^2-\epsilon^2]} .
\label{homoS}
\end{eqnarray}
%
After substituting these equations into Eqs.~(\ref{Iud2})--(\ref{Idd2}) and Eqs.~(\ref{SAR11})--(\ref{sigmaK})
and performing some algebraic transformations we obtain formulas (\ref{Iud3})--(\ref{Idd3}) which are discussed 
in detail in Section \ref{Sub}.
\end{appendix}


\end{multicols}
\end{document}